\documentclass[11pt]{article}
\pdfoutput=1

\usepackage{cancel,slashed}
\usepackage{bbm}
\usepackage{amssymb}
\usepackage{amsfonts}
\usepackage{amsmath}
\usepackage{graphicx}
\usepackage{cite}
\usepackage{physics}

\usepackage{ dsfont }
\usepackage{latexsym}
\usepackage{color}
\usepackage{cancel,slashed}
\usepackage[hyperfootnotes=false,linktocpage]{hyperref}
\usepackage{color}
\usepackage{transparent}
\usepackage[hang,flushmargin]{footmisc}

\usepackage{blkarray}
\usepackage{multirow}
\usepackage{empheq}

\usepackage{comment}

\newcommand{\bmat}{\left(\begin{array}}
\newcommand{\emat}{\end{array}\right)}

\newcommand{\pr}{\mathbbm{R}}

\def\Z{\mathbb{Z}}

\def\CK {{\cal K}}
\def\a {\alpha}
\def\b {\beta}

\def\ov{\overline}

\def\Tr{\text{Tr}}
\def\IM{\text{Im}\,}
\def\RE{\text{Re}\,}
\def\ov{\overline}
\def\1{{\bf 1}}
\def\2{{\bf 2}}
\def\3{{\bf 3}}
\def\4{{\bf 4}}
\def\6{{\bf 6}}

\def\targ#1#2{\genfrac{[}{]}{0pt}{}{#1}{#2}}
\def\targ2#1#2{\genfrac{}{}{0pt}{}{#1}{#2}}

\definecolor{mygr}{rgb}{0,0.6,0}
\definecolor{mygrey}{rgb}{0,0.1,0.2}
\definecolor{myblue}{rgb}{0,0.5,0.9}
\definecolor{myblue2}{rgb}{0,0.5,0.5}
\definecolor{myblue3}{rgb}{0,0.7,0.9}
\definecolor{myblue4}{rgb}{0,0.6,0.6}
\definecolor{myorange}{rgb}{1,0.5,0}
\definecolor{mypurple}{rgb}{0.6,0,1}
\definecolor{mygolden}{rgb}{1,0.8,0.2}
\definecolor{mycyan}{rgb}{0,1,1}
\definecolor{mymagenta}{rgb}{1,0,1}
\definecolor{mykiwi}{rgb}{0.8,1,0.5}
\definecolor{mybrown}{cmyk}{0.14, 0.42, 0.56, 0.2}
\definecolor{myturq}{cmyk}{0.99, 0, 0.2, 0.4}
\definecolor{myaubergine2}{cmyk}{0.4, 0.5, 0, 0.1}
\definecolor{myaubergine}{cmyk}{0.6,0.85,0,0}
\definecolor{CycleGreen}{cmyk}{0.52,0,1,0}
\definecolor{CycleBrown}{cmyk}{0, 0.4, 0.9, 0.2}

\DeclareFontFamily{U}{rcjhbltx}{}
\DeclareFontShape{U}{rcjhbltx}{m}{n}{<->rcjhbltx}{}
\DeclareSymbolFont{hebrewletters}{U}{rcjhbltx}{m}{n}

\DeclareMathSymbol{\lamed}{\mathord}{hebrewletters}{108}
\DeclareMathSymbol{\mem}{\mathord}{hebrewletters}{109}
\DeclareMathSymbol{\ayin}{\mathord}{hebrewletters}{96}
\DeclareMathSymbol{\tsadi}{\mathord}{hebrewletters}{118}
\DeclareMathSymbol{\qof}{\mathord}{hebrewletters}{113}
\DeclareMathSymbol{\resh}{\mathord}{hebrewletters}{114}
\DeclareMathSymbol{\pe}{\mathord}{hebrewletters}{112}
\DeclareMathSymbol{\pesofit}{\mathord}{hebrewletters}{80}
\DeclareMathSymbol{\samekh}{\mathord}{hebrewletters}{115}
\DeclareMathSymbol{\tav}{\mathord}{hebrewletters}{116}
\DeclareMathSymbol{\vav}{\mathord}{hebrewletters}{119}
\DeclareMathSymbol{\het}{\mathord}{hebrewletters}{120}
\DeclareMathSymbol{\yod}{\mathord}{hebrewletters}{121}
\DeclareMathSymbol{\zayin}{\mathord}{hebrewletters}{122}
\DeclareMathSymbol{\alephdot}{\mathord}{hebrewletters}{128}
\DeclareMathSymbol{\tsadisofit}{\mathord}{hebrewletters}{90}
\DeclareMathSymbol{\shin}{\mathord}{hebrewletters}{152}

\def\CN {{\cal N}}

\def\d{{\delta}}
\def\be{\begin{equation}}
\def\ee{\end{equation}}
\def\bea{\begin{eqnarray}}
\def\eea{\end{eqnarray}}
\def\bes{\begin{subequations}}
\def\ees{\end{subequations}}

\def\oh{\frac{1}{2}}

\def\re{\mbox{Re}\, }
\def\im{\mbox{Im}\, }

\def\cy {{\text{CY}}}
\def\IZ{\mathbb{Z}}

\def\Om{\Omega}
\usepackage{multicol}
\usepackage{float}
\usepackage{caption}
\def\p {{\partial}}

\def\CO {{\cal O}}

\newcommand{\cF}{\mathcal{F}}

\newcommand{\cK}{\mathcal{K}}

\newcommand{\cN}{\mathcal{N}}
\newcommand{\cO}{\mathcal{O}}

\newcommand{\II}{\mathbb{I}}

\newenvironment{eqn*}{\begin{equation*}\begin{aligned}}{\end{aligned}\end{equation*}\noindent}

\makeatletter
\newsavebox\myboxA
\newsavebox\myboxB
\newlength\mylenA

\newcommand*\xoverline[2][0.75]{%
\sbox{\myboxA}{$\m@th#2$}%
\setbox\myboxB\null% Phantom box
\ht\myboxB=\ht\myboxA%
\dp\myboxB=\dp\myboxA%
\wd\myboxB=#1\wd\myboxA% Scale phantom
\sbox\myboxB{$\m@th\overline{\copy\myboxB}$}%  Overlined phantom
\setlength\mylenA{\the\wd\myboxA}%   calc width diff
\addtolength\mylenA{-\the\wd\myboxB}%
\ifdim\wd\myboxB<\wd\myboxA%
   \rlap{\hskip 0.5\mylenA\usebox\myboxB}{\usebox\myboxA}%
\else
    \hskip -0.5\mylenA\rlap{\usebox\myboxA}{\hskip 0.5\mylenA\usebox\myboxB}%
\fi}
\makeatother

\begin{document}
\pagestyle{plain}

%----------------------------------------------------------------------%
%  numbering equations with section number
%----------------------------------------------------------------------%
\makeatletter
\@addtoreset{equation}{section}
\makeatother
\renewcommand{\theequation}{\thesection.\arabic{equation}}
%----------------------------------------------------------------------%
%  title page
%----------------------------------------------------------------------%

\pagestyle{empty}
%\vspace*{1.0in}
\rightline{IFT-UAM/CSIC-21-115}
%\rightline{\tt hep-th/yymmnnn}
\vspace{0.5cm}
\begin{center}
\Huge{{BIonic membranes \\ and  AdS instabilities} 
%\\ and AdS instabilities}
\\[10mm]}
\normalsize{Fernando Marchesano,  David Prieto, and Joan Quirant \\[12mm]}
\small{
Instituto de F\'{\i}sica Te\'orica UAM-CSIC, c/ Nicol\'as Cabrera 13-15, 28049 Madrid, Spain
\\[10mm]} 
\small{\bf Abstract} \\[5mm]
\end{center}
\begin{center}
\begin{minipage}[h]{15.0cm}

We study 4d membranes in type IIA flux compactifications of the form AdS$_4 \times X_6$, where $X_6$ admits a Calabi--Yau metric. These models feature scale separation and D6-branes/O6-planes on three-cycles of $X_6$. When the latter are treated as localised sources, explicit solutions to the 10d equations of motion and Bianchi identities are known in 4d $\cN=1$ settings, valid at first order in an expansion parameter related to the AdS$_4$ cosmological constant. We extend such solutions to a family of perturbatively stable $\cN=0$ vacua, and analyse their non-perturbative stability by looking at 4d membranes. Up to the accuracy of the solution, we find that either D4-branes or anti-D4-branes on holomorphic curves feel no force in both $\cN =1$ and $\cN=0$ AdS$_4$. Differently, D8-branes wrapping $X_6$ and with D6-branes ending on them can be superextremal 4d membranes attracted towards the $\cN=0$ AdS$_4$ boundary. The sources of imbalance are the curvature of $X_6$ and the D8/D6 BIon profile, with both comparable terms as can be checked for $X_6$ a (blown-up) toroidal orbifold.  We then show that simple $\cN=0$ vacua  with space-time filling D6-branes are unstable against bubble nucleation, decaying to $\cN=0$ vacua with less D6-branes and larger Romans mass.

\end{minipage}
\end{center}
\newpage
%----------------------------------------------------------------------%
%  Resetting of counters
%----------------------------------------------------------------------%
\setcounter{page}{1}
\pagestyle{plain}
\renewcommand{\thefootnote}{\arabic{footnote}}
\setcounter{footnote}{0}
%----------------------------------------------------------------------%
%  Paper begins
%----------------------------------------------------------------------%

%\end{document}

\tableofcontents

%%%%%%%%%%%%%%%%%%%
%%%%%%%%%%%%%%%%%%%
%\newpage
\section{Introduction}
\label{s:intro}

Out of the different aspects of the Swampland Programme \cite{Vafa:2005ui,Brennan:2017rbf,Palti:2019pca,vanBeest:2021lhn,Grana:2021zvf} one of the most far-reaching is the interplay between quantum gravity and supersymmetry breaking. In the specific context of non-supersymmetric vacua, several proposals for Swampland criteria put severe constraints on their stability. In particular, the AdS Instability Conjecture \cite{Ooguri:2016pdq,Freivogel:2016qwc} proposes that all $\cN=0$ AdS$_d$ vacua are at best metastable, with  bubble nucleation always mediating some non-perturbative decay. The motivation for this proposal partially arises from a refinement of the Weak Gravity Conjecture (WGC) stating that the WGC inequality is only saturated in supersymmetric theories \cite{Ooguri:2016pdq}. Applied to $(d-2)$-branes, this implies a specific decay mechanism for $\cN=0$ AdS$_d$ vacua supported by $d$-form background fluxes, in which a probe superextremal $(d-2)$-brane nucleates and expands towards the AdS$_d$ boundary, as in \cite{Maldacena:1998uz}.

These proposals have been tested in different contexts, and in particular for type II setups in which the AdS solution is supported by fluxes \cite{Gaiotto:2009mv,Antonelli:2019nar,Apruzzi:2019ecr,Bena:2020xxb,Suh:2020rma,Guarino:2020jwv,Guarino:2020flh,Basile:2021vxh,Apruzzi:2021nle,Bomans:2021ara}. Remarkably, compactifications of the form AdS$_4 \times X_6$, where $X_6$ admits a Calabi--Yau metric \cite{Villadoro:2005cu,DeWolfe:2005uu,Camara:2005dc} remain elusive of the conjecture, because so far  the decays  observed for perturbatively stable $\cN=0$ vacua are marginal \cite{Narayan:2010em}, and the corresponding membranes saturate the WGC inequality. A better understanding of these constructions seems thus crucial to the Swampland Programme: Their non-supersymmetric version challenges the AdS Instability Conjecture, and more precisely the WGC for membranes, while the supersymmetric settings challenge the strong version of the AdS Distance Conjecture \cite{Lust:2019zwm}. As pointed out in \cite{Buratti:2020kda}, the tension with the AdS Distance Conjecture could be solved by taking into account the discrete symmetries related to 4d membranes, so the spectrum and  properties of 4d membranes seem to be at the core of both issues.  Finally, the constructions in \cite{DeWolfe:2005uu,Camara:2005dc} are particularly interesting phenomenologically, since besides supersymmetry breaking they incorporate key features like scale separation and chiral gauge theories supported on D6-branes wrapping intersecting three-cycles of $X_6$.

An important caveat of the constructions in \cite{Villadoro:2005cu,DeWolfe:2005uu} is that they do not solve the 10d equations of motion and Bianchi identities, unless localised sources like D6-branes and O6-planes are smeared over the internal dimensions \cite{Acharya:2006ne}. Nevertheless, one may look for solutions with localised sources by formulating the problem as a perturbative expansion, of which the leading term is the smeared-source Calabi--Yau approximation, and where the expansion parameter is essentially the AdS$_4$ cosmological constant \cite{Saracco:2012wc}. The first-order correction to the smeared background was found in \cite{Junghans:2020acz,Marchesano:2020qvg}, displaying localised sources and a natural expansion parameter $R^{-4/3} \sim g_s^{4/3}$, where $R$ is the AdS$_4$ radius in string units and $g_s$ is the average 10d string coupling.\footnote{Another caveat surrounding these constructions is that they combine O6-planes and a non-vanishing Romans mass, which makes difficult to understand them microscopically. However, T-dual versions of the solutions in \cite{Junghans:2020acz,Marchesano:2020qvg} have been constructed in \cite{Cribiori:2021djm} with similar properties, vanishing Romans mass and an 11d description.}  

In this paper we revisit the stability of the AdS$_4$ vacua in \cite{DeWolfe:2005uu,Camara:2005dc,Narayan:2010em}, with the vantage point of the more precise 10d description of \cite{Junghans:2020acz,Marchesano:2020qvg}. We consider $\cN=1$ and $\cN=0$ vacua which, in the smearing approximation,  are related by an overall sign flip of the internal four-form flux $G_4$. These were considered in \cite{DeWolfe:2005uu,Narayan:2010em} for $X_6 = T^6/ (\Z_3 \times \Z_3)$, and generalised to arbitrary Calabi--Yau geometries in \cite{Marchesano:2019hfb}. For the supersymmetric backgrounds a rather explicit 10d solution was provided in \cite{Marchesano:2020qvg}, in terms of an $SU(3)\times SU(3)$-structure deformation of the Calabi--Yau metric. For their non-supersymmetric cousins we use the approach in \cite{Junghans:2020acz} to provide a solution at the same level of approximation. 
In this setup we consider 4d membranes that come from wrapping D$(2p)$-branes on $(2p-2)$-cycles of $X_6$. These membranes couple to  fluxes that support the AdS$_4$ background, more precisely to the dynamical fluxes of the 4d theory \cite{Lanza:2019xxg,Lanza:2020qmt}. Therefore, even if there could be other non-perturbative decay channels, the $\cN=0$ sharpening of the WGC suggests that at least one of these membranes or a bound state of them should be superextremal, and thus a candidate to yield an expanding bubble. Note that these AdS$_4$ backgrounds have  not been constructed as near-horizon limits of a backreacted black brane solutions, so it is a priori not clear which membrane is the most obvious candidate to fulfil the conjecture. 

It was argued in \cite{Aharony:2008wz,Narayan:2010em} that D4-branes wrapping either holomorphic or anti-holomorphic cycles of $X_6$ saturate a BPS bound for the $\cN=1$ and $\cN=0$ vacua mentioned above, while D2-branes and D6-branes wrapping four-cycles never do. By looking at each of their couplings to the fluxes supporting the AdS$_4$ background and their tension we recover the same result. Remarkably, we not only do so for the smeared-source Calabi--Yau  approximation considered in \cite{Aharony:2008wz,Narayan:2010em}, but also when the first-order corrections to this background are taken into account. It follows that at this level of approximation such (anti-)D4-branes give rise to extremal objects, that can at most mediate marginal decays. This extends to bound states of D6, D4 and D2-branes, in the sense that they do not yield any superextremal 4d membrane. 

We then turn to consider D8-branes wrapping $X_6$. Due to a Freed--Witten anomaly generated by the $H$-flux, D6-branes must be attached to the D8 worldvolume. From the 4d perspective, these are membranes that not only change the Romans mass flux $F_0$ when crossing them, but also the number of space-time filling D6-branes, so that the tadpole condition is still satisfied. It turns out that the presence of attached D6-branes acts as a force on the D8-branes, and exactly cancels the effect of their charge and tension in supersymmetric vacua, as it should happen for a BPS object. This provides a rationale  for the precise relation between $F_0$, $R$, $g_s$ found in \cite{DeWolfe:2005uu}. In $\cN=0$ vacua the energetics of D8-branes is more interesting, because curvature corrections induce D4-brane charge and tension on their worldvolume. The induced tension is in general negative, implying that the D8-brane is dragged towards the boundary of $\cN=0$ AdS$_4$.  As we argue, this corresponds to a superextremal 4d membrane that mediates a decay to another non-supersymmetric vacuum with larger $|F_0|$ and fewer D6-branes, in agreement with the sharpened Weak Gravity Conjecture.

This picture is however incomplete, since it relies on the smeared description. First-order corrections to the Calabi--Yau background modify the D8-brane action by terms comparable to an induced D4-brane tension. In fact, beyond the smearing approximation the D8/D6 system should be treated as a BIon-like solution, whose tension differs from the sum of D8 and D6-brane tensions. We compute this difference for $X_6 = T^6/(\Z_2 \times \Z_2)$, and find that this new correction is comparable to curvature-induced effects. Nevertheless, for simple D-brane configurations we find that it is also negative, and so the D8-branes are still dragged towards the $\cN=0$ AdS$_4$ boundary. If the same is true in more general setups, then the combined effect of curvature and BIon-like corrections provide a non-perturbative instability for $\cN=0$ AdS$_4$ vacua with space-time filling D6-branes, in line with the AdS Instability Conjecture.

The paper is organised as follows. In section \ref{s:memb} we discuss the energetics of membranes in AdS$_4$ backgrounds with four-form fluxes, which we then use as a criterion for membrane extremality. In section \ref{s:dgkt} we review the $\cN=1$ AdS$_4$ Calabi--Yau orientifold vacua with fluxes in the smearing approximation, and classify BPS membranes that come from wrapped D-branes. Section \ref{s:nonsusy} does the same for non-supersymmetric AdS$_4$, finding superextremal membranes thanks to curvature corrections, and section \ref{s:insta} argues that they mediate actual decays in the 4d theory. Section \ref{s:nonsmeared} describes the 10d background with localised sources for $\cN=1$ and $\cN=0$ AdS$_4$ vacua, and shows that D4-branes saturate a BPS bound in both cases. Section \ref{s:bion} describes D8/D6-brane systems as BIons, and shows that they are BPS in $\cN=1$ but feel a net force in $\cN=0$ vacua. We finally present our conclusions in section \ref{s:conclu}.

Several technical details have been relegated to the appendices. Appendix \ref{ap:CYIIA} reviews type IIA moduli stabilisation on Calabi--Yau with fluxes, adapted to our conventions. Appendix \ref{ap:10deom} shows that the  backgrounds of section \ref{s:nonsmeared} satisfy the 10d equations of motion. Appendix \ref{ap:dbi} shows how the BIon profile of section \ref{s:bion} linearises the  DBI action. Appendix \ref{ap:IIBion} relates this profile to 4d strings in type IIB warped Calabi--Yau compactifications and to SU(4) instantons in Calabi--Yau four-folds. Appendix \ref{ap:torus} computes the  BIonic D8-branes excess tension for a simple D-brane configuration in $X_6 = T^6/(\Z_2 \times \Z_2)$.

%%%%%%%%%%%%%%%%%%%
%%%%%%%%%%%%%%%%%%%

\section{Membranes in AdS$_4$}
\label{s:memb}

In a 4d Minkowski background with $\cN=1$ supersymmetry, simple examples of static BPS membranes are  3d hyperplanes of $\pr^{1,3}$ including the time-like direction. Analogous objects in anti-de Sitter can be described by considering the Poincar\'e patch of AdS$_4$, whose metric reads
\be
ds^2_4 =e^{\frac{2z}{R}} (-dt^2 + d\vec{x}^2) + dz^2\, ,
\label{PPatch}
\ee
with $R$ the AdS length scale, $\vec{x} = (x^1, x^2)$, and all coordinates range over $\pr$. In such coordinates, the AdS$_4$ boundary is located at $z = \infty$. Similarly to the Minkowski case, one may consider a membrane that spans the coordinate $t$ and a surface within $(x^1, x^2, z)$. Particularly simple is the case where the surface is the plane $z=z_0$, with $z_0 \in \pr$ fixed. While this object may look like the BPS membranes of Minkowski, the tension of such a membrane decreases exponentially as we take $z_0 \to -\infty$. Therefore, if we place such an object in AdS$_4$ and take the probe approximation, it will inevitably be driven away from the boundary and it cannot be BPS. 

This can be avoided if on top of the AdS$_4$ metric we consider a four-form flux background $F_4$, to whose three-form potential $C_3$ the membrane couples as $-\int C_3$. Indeed, if we have
\be
\langle F_4 \rangle = -\frac{3Q}{R} {\rm vol}_4
\qquad \Longrightarrow \qquad \langle C_3 \rangle = Q\, e^{\frac{3z}{R}} dt \wedge dx^1 \wedge dx^2 \, ,
\label{3form}
\ee
and $Q$ coincides with the tension of the membrane $T$, then the variation of the tension when moving in the $z$ coordinate is compensated by the potential energy $-\int \langle C_3 \rangle$ gained because of its charge. Moving along this coordinate is then a flat direction and the membrane may be BPS. If $Q > T$ one may still find BPS membrane configurations, but they cannot be parallel to the boundary. We instead have that force cancellation occurs for embeddings of the form
\be
\left\{t, x^1, x^2 = \pm \frac{R}{\sqrt{\frac{Q^2}{T^2}-1}} \, e^{-\frac{z}{R}} + c\right\}\, , \qquad c \in \pr\, .
 \label{QnotT}
\ee

Four-form flux backgrounds are ubiquitous in AdS$_4$ backgrounds obtained from string theory, and in particular in those with 4d $\cN=1$ supersymmetry or $\cN=0$ spontaneously broken. The membrane profiles $z=z_0$ and \eqref{QnotT} were found in \cite{Koerber:2007jb} in the context of $\cN=1$ AdS$_4$ backgrounds obtained from type II string theory, but from the above discussion it follows that they can also be present in backgrounds with supersymmetry spontaneously broken by fluxes.  One can in fact see that  the set of 4d fluxes arising from the compactification is directly related to the spectrum of BPS branes, as well as to the internal data specifying the supersymmetry generators. 

In the following we will be chiefly concerned with those membranes whose profile is given by $z=z_0$. As argued in \cite{Koerber:2007jb}, for $Q=T$ and at $z \to \infty$ they capture the BPS bound of a spherical membrane in global coordinates at asymptotically large radius. It is precisely the domain walls that correspond to spherical membranes near the AdS boundary that determine if the non-perturbative decay of one vacuum to another with lower energy is favourable or not. Thus, by considering the energetics of membranes in the Poincar\'e patch with $z = z_0 \to \infty$ we may detect if there could be some domain wall triggering such a decay. If all membranes satisfy $T > Q$ such a decay should not occur, if $Q=T$ it should be marginal, and  if $T<Q$ the AdS background may develop a non-perturbative instability. According to the conjectures in \cite{Ooguri:2016pdq,Freivogel:2016qwc}, any $\cN=0$ AdS background of this sort should have at least one non-perturbative instability towards a new vacuum, and therefore a membrane with $T<Q$. In the following sections we will consider the membranes that appear from wrapping D-branes on internal cycles in backgrounds of the form AdS$_4 \times X_6$, where $X_6$ admits a Calabi--Yau metric, and compute $T$ and $Q$ for them. In particular we will consider the $\cN=1$ vacua of \cite{DeWolfe:2005uu} and some of the non-supersymmetric vacua found in \cite{Camara:2005dc,Narayan:2010em,Marchesano:2019hfb}, which are stable at the perturbative level. We will not only consider the Calabi--Yau approximation of these references, but also the solutions with localised sources found in \cite{Junghans:2020acz,Marchesano:2020qvg}. As we will see, for non-supersymmetric vacua the answer is not the same once this more precise picture is taken into account.

%%%%%%%%%%%%%%%%%%%
%%%%%%%%%%%%%%%%%%%

\section{Supersymmetric AdS$_4$ orientifold vacua}
\label{s:dgkt}

Examples of membranes satisfying $Q=T$ are typically found in supersymmetric AdS$_4$ backgrounds, where the equality follows from saturating a BPS bound. In this section we analyse for which membranes this condition is met for the supersymmetric type IIA flux compactifications of \cite{DeWolfe:2005uu}, for an arbitrary Calabi--Yau geometry $X_6$, in the approximation of smeared sources \cite{Acharya:2006ne}. With the simple criterion $Q=T$ one can reproduce the results of \cite{Aharony:2008wz} for membranes arising from D2, D4 and D6-branes wrapping internal cycles of $X_6 = T^6/(\Z_3 \times \Z_3)$, and extend them to any Calabi--Yau manifold. Furthermore, one may detect an additional set of BPS membranes, namely those coming from D8-branes wrapping $X_6$, to which space-time filling D6-branes are attached. This last feature makes such membranes quite special, particularly when one considers them for non-supersymmetric AdS$_4$ backgrounds and beyond the smearing approximation.

\subsection{10d background in the smearing approximation}
\label{ss:smeared}

Let us consider type IIA string theory compactified in an orientifold of $X_4 \times X_6$, where $X_6$ is a compact Calabi--Yau three-fold. As in the standard construction \cite{Ibanez:2012zz}, we take the orientifold action to be generated by $\Omega_p (-1)^{F_L}{\cal R}$,\footnote{Here $\Omega_p$ is the worldsheet parity reversal operator, ${F_L}$ is the space-time fermion number for the left-movers.} with ${\cal R}$ an anti-holomorphic involution of $X_6$ acting as ${\cal R} J_{\rm CY}=-J_{\rm CY}$ and ${\cal R}\Omega_{\rm CY} = - \overline{\Omega}_{\rm CY}$ on its K\"ahler 2-form and holomorphic 3-form, respectively. The fixed locus $\Pi_{\rm O6}$ of ${\cal R}$ is one or several 3-cycles of $X_6$ in which O6-planes are located. Further localised sources may include D6-branes wrapping three-cycles and coisotropic D8-branes \cite{Font:2006na}. Together with the contribution of background fluxes they must cancel the O6-plane RR charge.

One convenient way to describe the background fluxes of the compactification is to use the democratic formulation of type IIA supergravity \cite{Bergshoeff:2001pv},  in which all RR potentials are grouped in a polyform ${\bf C} = C_1 + C_3 + C_5 + C_7 + C_9$, and so are their gauge invariant field strengths
\be
{\bf G} \,=\, d_H{\bf C} + e^{B} \wedge {\bf \bar{G}} \, ,
\label{bfG}
\ee
with $H$ the three-form NS flux, $d_H \equiv (d - H \wedge)$ is the $H$-twisted differential  and ${\bf \bar{G}}$ a formal sum of closed $p$-forms on $X_6$. The Bianchi identities read
\begin{equation}\label{IIABI}
\ell_s^{2} \,  d (e^{-B} \wedge {\bf G} ) = - \sum_\a \lambda \left[\delta (\Pi_\alpha)\right] \wedge e^{\frac{\ell_s^2}{2\pi} F_\alpha} \, ,  \qquad d H = 0 \, ,
\end{equation} 
with $\ell_s  =  2\pi \sqrt{\a'}$ the string length. Here $\Pi_\alpha$ hosts a D-brane source with a quantised worldvolume flux $F_\alpha$, and $\delta(\Pi_\alpha)$ is the bump $\delta$-function form with support on $\Pi_\alpha$ and indices transverse to it, such that $\ell_s^{p-9} \d(\Pi_\a)$ lies in the Poincar\'e dual class to $[\Pi_\a]$. O6-planes contribute as D6-branes but with minus four times their charge and $F_\alpha \equiv 0$. Finally, $\lambda$ is the operator that reverses the order of the indices of a $p$-form.

 In the absence of localised sources, each $p$-form within ${\bf \bar{G}}$ is quantised, so one can define the internal RR flux quanta in terms of the following integer numbers
\begin{equation}
m \, = \,  \ell_s G_0\, ,  \quad  m^a\, =\, \frac{1}{\ell_s^5} \int_{X_6} \bar{G}_2 \wedge \tilde \omega^a\, , \quad  e_a\, =\, - \frac{1}{\ell_s^5} \int_{X_6} \bar{G}_4 \wedge \omega_a \, , \quad e_0 \, =\, - \frac{1}{\ell_s^5} \int_{X_6} \bar{G}_6 \, ,
\label{RRfluxes}
\end{equation}
with $\omega_a$, $\tilde \omega^a$ integral harmonic two- and four-forms such that $\ell_s^{-6} \int_{X_6} \omega_a \wedge \tilde{\omega}^b = \delta_a^b$, in terms of which we can expand the K\"ahler form as
\be
J_{\rm CY} = t^a \omega_a\, , \qquad - J_{\rm CY} \wedge J_{\rm CY} = {\cal K}_a \tilde{\omega}^a\, . 
\ee
Here ${\cal K}_a \equiv {\cal K}_{abc} t^bt^c$, with ${\cal K}_{abc} = - \ell_s^{-6} \int_{X_6} \omega_a \wedge \omega_b \wedge \omega_c$ the Calabi--Yau triple intersection numbers and $-\frac{1}{6} J_{\rm CY}^3 = - \frac{i}{8} \Omega_{\rm CY} \wedge \bar{\Omega}_{\rm CY}$ its volume form.\footnote{Due to this choice of volume form the triple intersection numbers must be defined with an additional minus sign compared to the more standard definition in the literature so that, whenever $\{[\ell_s^{-2}\omega_a]\}_a$ is dual to a basis of Nef divisors, ${\cal K}_{abc} \geq 0$. The same observation applies to the curvature correction term $K_a^{(2)}$ defined in \eqref{Kcurv}.}

In the presence of D6-branes and O6-planes the Bianchi identities for the RR fluxes read
\be
dG_0 = 0\, , \qquad d G_2 = G_0 H - 4 \d_{\rm O6} +   N_\a \d_{\rm D6}^\a \, ,  \qquad d G_4 = G_2 \wedge H\, , \qquad dG_6 = 0\, ,
\label{BIG}
\ee
where  we have defined $\d_{\rm D6/O6}\equiv \ell_s^{-2}  \d(\Pi_{\rm D6/O6})$. This in particular implies that
\be
{\rm P.D.} \left[4\Pi_{\rm O6}- N_\a \Pi_{\rm D6}^\a\right] = m [\ell_s^{-2} H] \, ,
\label{tadpole}
\ee
constraining the quanta of Romans parameter and NS flux. Let us in particular choose P.D.$[\ell_s^{-2}H] = h [\Pi_{\rm O6}] = h [\Pi_{\rm D6}^\a]$, $\forall \a$. We then find the constraint
\be
mh +N = 4\, ,
\label{tadpole2}
\ee
with $N$ the number of D6-branes wrapping $\Pi_{\rm O6}$. Supersymmetry in addition implies that $mh$ and $N$ are non-negative, yielding a finite number of solutions.\footnote{In several instances (e.g., toroidal orbifolds) $[\Pi_{\rm O6}]$ may be an integer multiple $k$ of a three-cycle class. In those cases $h, N$ need not be integers, but instead $kh, kN \in \mathbb{Z}$, allowing for a richer set of solutions to \eqref{tadpole2}.} 

The constraint on sign$(mh)$ can be seen by means of a 4d analysis of the potential generated by background fluxes, following \cite{DeWolfe:2005uu,Camara:2005dc}. Such a potential was obtained in \cite{Grimm:2004ua} by combining the superpotential generated by the RR and NS flux quanta and the classical K\"ahler potential of Calabi--Yau orientifolds without fluxes. We review such a 4d analysis, from the more general viewpoint of \cite{Marchesano:2019hfb}, in Appendix \ref{ap:CYIIA}. One of the main results is an infinite discretum of $\cN=1$ AdS$_4$ vacua, associated to an internal Calabi--Yau manifold $X_6$ such that the internal fluxes satisfy
\be
\ell_s [ H ]  = \frac{2}{5} m g_s [\re \Omega_{\rm CY} ] \, , \qquad  G_2  =  0\, ,  \qquad  \ell_s G_4  = -\hat{e}_a  \tilde{\omega}^a  = -\frac{3}{10} m \, \cK_a \tilde{\omega}^a  \, , \qquad  G_6 =  0\, , 
\label{intfluxsm}
\ee
where we have defined
\be
\hat{e}_a = e_a - \oh \frac{\cK_{abc} m^bm^c}{m}\, .
\label{hate}
\ee

Care should however be taken when interpreting such relations from the viewpoint of the actual 10d supergravity solution, since the presence of fluxes and localised sources will deform the internal geometry away from the Calabi--Yau metric, and a $G_2$ and $G_4$ of the above form will never satisfy the Bianchi identities \eqref{BIG}.\footnote{Additionally, in the presence of D6-brane moduli the integral of $\bar{G}_2$ will depend on them. This can  be dealt with by translating such a dependence into a superpotential involving both open and closed string moduli \cite{Marchesano:2014iea,Carta:2016ynn}.} The standard way to deal with both issues is to see \eqref{intfluxsm} as a formal solution in which all localised sources have been smeared \cite{Acharya:2006ne}. This so-called smeared solution is then the leading term in a perturbative series that should converge to the actual background \cite{Saracco:2012wc}, with expansion parameter $g_s^{4/3}$, and where sources are localised \cite{Junghans:2020acz,Marchesano:2020qvg}.  Instead of \eqref{intfluxsm}, the relations that this background must satisfy are
\be
[ H ]  = \frac{2}{5} G_0 g_s  [\re \Omega_{\rm CY} ] \, , \quad \int_{X_6} G_2 \wedge \tilde{\omega}^a =  0\, ,  \quad \frac{1}{\ell_s^6} \int_{X_6} G_4  \wedge \omega_a  =  - \frac{3}{10} G_0 {\cal K}_a \, , \quad  G_6  =  0\, , 
\label{intflux}
\ee
where $g_s$ is the average value of $e^{\phi}$, with $\phi$ a varying 10d dilaton. This value determines the AdS$_4$ length scale in the 10d string  frame $R$, from the following additional relation
\be
\frac{\ell_s}{R} = \frac{1}{5} |m| g_s  \, .
\label{Rads}
\ee
There is in addition a non-trivial warp factor, and the Calabi--Yau metric on $X_6$ is deformed to an $SU(3) \times SU(3)$-structure metric \cite{Marchesano:2020qvg}. We will discuss this more accurate background in section \ref{ss:mpqt}, and for now focus on the smearing approximation. 

It follows from such a description that the Calabi--Yau volume 
\be
{\cal V}_{\rm CY}  = -\frac{1}{6\ell_s^6} \int_{X_6} J_{\rm CY}^3 =  \frac{1}{6} {\cal K}_{abc}t^at^bt^c \equiv  \frac{1}{6} {\cal K}\, ,
\ee
depends on $m$ and $\hat{e}_a$, growing larger when we increase their absolute value. One can then for instance see that $1/R$ grows as we increase $h$ or $m$, and decreases as we increase $\hat{e}_a$ The more precise result can be obtained from the 4d analysis, which yields the following 4d Einstein frame vacuum energy 
\be
\Lambda = - \frac{16\pi}{75\kappa_4^4} e^K {\cal K}^2 m^2 \, ,
\label{lambda}
\ee
where $K$ is the K\"ahler potential, given by \eqref{WK}. One can then see that $\Lambda$ scales like $|m|^{5/2}$, as in the explicit toroidal solutions in \cite{DeWolfe:2005uu,Camara:2005dc}. Recall however that the allowed values for $m$ are bounded by the tadpole condition \eqref{tadpole2}.

Finally, one can include the effect of curvature corrections to the 4d analysis, following \cite{Palti:2008mg,Escobar:2018rna}. We will only include those corrections dubbed $K^{(1)}_{ab}$ and $K_a^{(2)}$ in \cite{Palti:2008mg,Escobar:2018rna}, given by
\be
K^{(1)}_{ab} = \frac{1}{2} {\cal K}_{aab} \, , \qquad K_a^{(2)} = -\frac{1}{24\ell_s^6} \int_{X_6} c_2(X_6) \wedge \omega_a\, ,
\label{Kcurv}
\ee
which respectively correspond to $\cO(\alpha')$ and $\cO(\alpha'^2)$ corrections, since higher orders will be beyond the level of accuracy of our analysis. If $\{[\ell_s^{-2}\omega_a]\}_a$ is dual to a basis of Nef divisors, then $K_a^{(2)} \geq 0$ \cite{Miyaoka1987}. The effect of such corrections is to redefine the background flux quanta as follows
\be
e_0 \to e_0 - m^a K_a^{(2)} \, , \qquad \quad e_a \to e_a  - K_{ab}^{(1)} m^b + m K_a^{(2)} \, ,
\label{curvflux}
\ee 
so in particular they modify the flux combinations \eqref{hate} that determine the K\"ahler moduli vevs. This modification makes more involved the scaling of $\Lambda$ with $m$, but since in the regime of validity we have that ${\cal K}_a \gg K_a^{(2)}$, it turns out that $\Lambda \sim |m|^{5/2}$ is still a good approximation. 

\subsection{4d BPS membranes}
\label{ss:4dmem}

Given a type II flux compactification to $\CN=1$ AdS$_4$, one may study the spectrum of BPS D-branes via $\kappa$-symmetry or pure spinor techniques, as in \cite{Aharony:2008wz,Koerber:2007jb}, and in particular determine those D-branes that give rise to BPS membranes from the 4d perspective. In the following we will take the more pedestrian viewpoint of section \ref{s:memb} to identify such BPS membranes. This criterion will also be useful when considering non-supersymmetric AdS$_4$ vacua. 

An analysis of 4d BPS membranes parallel to the AdS$_4$ boundary in the Poincar\'e patch was carried out in \cite{Aharony:2008wz}, for the particular case $X_6 = \mathbb{T}^6/(\Z_3 \times \Z_3)$ of \cite{DeWolfe:2005uu}, in the smearing approximation. It was found that D4-branes wrapping holomorphic cycles are BPS, while D2 and D6 branes cannot be so. Let us see how to recover such results and extend them to general Calabi--Yau geometries using the picture of section \ref{s:memb}. For this we recall that in the type IIA democratic formulation the RR background fluxes take the form
\be
{\bf G} = {\rm vol}_4 \wedge \tilde{G} + \hat{G}\, ,
\label{demoflux}
\ee
where ${\rm vol}_4$ is the AdS$_4$ volume form and $\tilde{G}$ and $\hat{G}$ only have internal indices, satisfying the relation $\tilde{G} = - \lambda ( *_6 \hat{G})$. Therefore from \eqref{intfluxsm} and \eqref{Rads} we find the following fluxes that translate into a 4d four-form background 
\be
G_6 =  - \frac{3\eta}{Rg_s} {\rm vol}_4 \wedge J_{\rm CY}\, , \qquad G_{10} =  - \frac{5\eta}{6 R g_s} {\rm vol}_4 \wedge J^3_{\rm CY}\, ,
\label{610fluxes}
\ee
with $\eta =  {\rm sign }\, m$. Contrarily, no component of ${\rm vol}_4$ appears in $G_4$ or $G_8$. We hence deduce the following  couplings for 4d membranes arising from D(2$p$)-branes wrapping (2$p-$2)-cycles of $X_6$:
\be
Q_{\rm D2} = 0 \, , \qquad Q_{\rm D4} =  e^{K/2} \frac{\eta}{\ell_s^2} \int_\Sigma J_{\rm CY} \, , \qquad Q_{\rm D6} = 0\, , \qquad Q_{\rm D8} =  -\frac{5}{3} e^{K/2} \eta\, q_{\rm D8} {\cal V}_{\rm CY}\, ,
\label{QDGKT}
\ee
expressed in 4d Planck units. Here $\Sigma$ is the two-cycle wrapped by the D4-brane, and $q_{\rm D8} = \pm 1$ specifies the orientation with which the D8-brane wraps $X_6$. This implies that for $\eta=1$ a BPS D4-brane must wrap a holomorphic two-cycle with vanishing worldvolume flux $\cF = B + \frac{\ell_s^2}{2\pi} F$ to be BPS, so that $e^{K/2} \ell_s^{-2}\int_\Sigma J_{\rm CY} = e^{K/2} {\rm area}(\Sigma)/\ell_s^2 \equiv T_{\rm D4}$, while for $\eta = -1$ the two-cycle must be anti-holomorphic. This choice of orientation for $\Sigma$ can be understood from looking at how the four-form varies when crossing the D4-brane from $z = \infty$ to $z = -\infty$. In both cases, due to \eqref{IIABI} and the choice of orientation for $\Sigma$ one decreases the absolute value of the four-form flux quanta $\hat{e}_a$, and therefore the vacuum energy. This is consistent with our expectations, as it permits to have a BPS domain-wall solution mediating a marginal decay from a vacuum with higher energy (at $z = \infty$) to one with lower energy (at $z = -\infty$). Considering this set of BPS membranes allows us to scan over the set of vacua with different four-form flux quanta.  Differently, D6-branes wrapping four-cycles of $X_6$ and D2-branes can never yield 4d BPS membranes. This indeed reproduces and generalises the results found in \cite{Aharony:2008wz}, adapted to our conventions. 

It however remains to understand the meaning of $Q_{\rm D8}$, which naively does not seem to allow for BPS membranes that come from wrapping (anti-)D8-branes on $X_6$. On general grounds one would expect that such BPS membranes exist as well, in order to scan over the different values of $m$. In particular, one would expect that for $\eta=1$ D8-branes $(q_{\rm D8} =1)$ wrapping $X_6$ are BPS, while for $\eta=-1$ the same occurs for anti-D8-branes $(q_{\rm D8} = -1)$. Indeed, when crossing the corresponding domain wall from $z = \infty$ to $z = -\infty$ the value of $|m|$ increases and the vacuum energy decreases in both setups, paralleling the case for D4-branes. However, the factor of $5/3$ and a sign prevent achieving the necessary BPSness condition $Q_{\rm D8}=T_{\rm D8} \equiv e^{K/2} {\cal V}_{\rm CY}$.

\begin{figure}[h!]
    \centering
    \includegraphics[width=10cm]{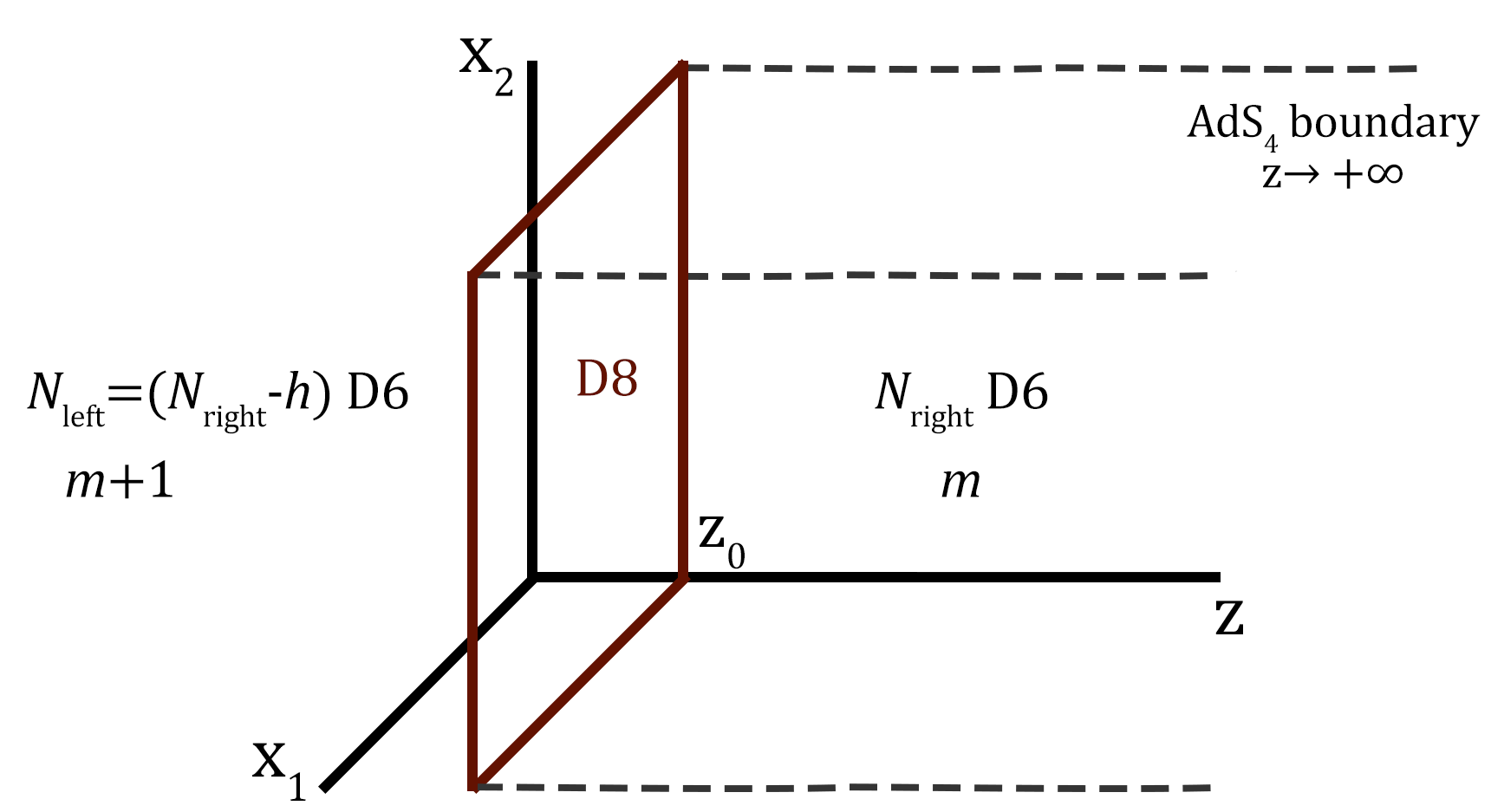}
    \caption{To cure the Freed--Witten anomaly induced by the $H$-flux on the D8-brane worldvolume, an excess of $|h|$ space-time filling D6-branes must be attached from its position to the AdS$_4$ boundary. We take $m, h >0$ in the figure.}
    \label{fig:D8D6}
\end{figure}

The resolution to this puzzle comes from realising that D8-branes wrapping $X_6$ cannot be seen as isolated objects. Instead, D6-branes must be attached to them, to cure the Freed--Witten anomaly generated on the (anti-)D8-brane by the NS flux background $H$. In the above setup the D6-branes will be wrapping a three-cycle of $X_3$ on the Poincar\'e dual class to $\eta [\ell_s^{-2} H] = |h| [\Pi_{\rm O6}]$, and extend along the 4d region of AdS$_4$ $(t,x^1,x^2) \times [z_0 , \infty)$ that is bounded by the 4d membrane. More generally, we need an excess of space-time filling D6-branes wrapping $\Pi_{\rm O6}$ on the interval $[z_0 , \infty)$ to the right of the (anti-)D8-brane, as compared to the ones in the left-interval $(-\infty, z_0]$ to cancel the said Freed--Witten anomaly: 
\be
N_{\rm right} - N_{\rm left} = |h| \, ,
\label{excessD6}
\ee
see figure \ref{fig:D8D6}. Since $m$ jumps by $\eta$ when crossing the membrane from right to left, $mh$ jumps by $|h|$, and so \eqref{excessD6} guarantees that the tadpole condition \eqref{tadpole2} is satisfied at both sides. 

Since the number of space-time filling D6-branes is different at both sides of the D8-brane, their presence will induce an energy dependence in terms of the D8-brane position. Indeed, if we decrease $z_0$ and move the D8-brane away from the AdS$_4$ boundary the region of AdS$_4$ filled by $N_{\rm right}$ D6-branes will grow, and so will the total energy of the system. As a result, the D6-brane jump induced by the Freed--Witten anomaly pulls the D8-branes towards the boundary of AdS$_4$. It turns out that this effects precisely cancels the effect of the tension $T_{\rm D8}$ and coupling $Q_{\rm D8}$ of the D8-brane, which both drag the 4d membrane away from the AdS boundary. 

One can derive such a cancellation via a microscopic calculation of the DBI+CS action for the D8/D6 system, dimensionally reduced to 4d. Of course, from the viewpoint of the 4d membrane the tension of space-time filling D6-branes extended along $(-\infty, z_0]$ and $[z_0, \infty)$ is infinite. Nevertheless, one may compute how the energy of the system varies  as we modify the D8-brane position $z_0$. Indeed, the DBI contribution to the action is given by the sum of the following two terms:
\begin{align}
 \label{DBID8}
    S_{\rm DBI}^{\rm D8} =& -\frac{1}{g_s} {\cal V}_{\rm CY}e^{\frac{3z_0}{R}} \frac{2\pi}{\ell_s^3} \int  dt dx^1 dx^2\, , \\
    S_{\rm DBI}^{\rm D6} = &
    -\frac{1}{g_s}  {\cal V}_{\Pi_{\rm O6}} \frac{2\pi}{\ell_s^4} \left(N_{\rm left} \int_{-\infty}^{z_0} dz'e^{\frac{3z'}{R}} +N_{\rm right} \int_{z_0}^\infty dz'e^{\frac{3z'}{R}}\right) \int  dt dx^1 dx^2\, ,
    \label{DBID6}
    \end{align}
with
\be
{\cal V}_{\Pi_{\rm O6}} = \frac{1}{\ell_s^3} \int_{\Pi_{\rm O6}} \im\, \Omega_{\rm CY} = \frac{1}{h \ell_s^5}  \int_{X_6}  \im \Omega_{\rm CY} \wedge H  = \frac{8}{5} \frac{m}{h} g_s {\cal V}_{\rm CY} = \frac{8\ell_s}{|h|R}{\cal V}_{\rm CY} \, ,
\ee
where we have used that in our conventions O6-planes and BPS D6-branes are calibrated by $\im\, \Omega_{\rm CY}$, and then the relations \eqref{intflux} and \eqref{Rads}. Let us now consider an infinitesimal variation $z_0 \to z_0 + \ell_s \epsilon$.  The variation of these actions is
\begin{align}
\label{DBID8var}
\delta_\epsilon  S_{\rm DBI}^{\rm D8} & = -\frac{3}{Rg_s } {\cal V}_{\rm CY}\, e^{\frac{3z_0}{R}}   \frac{2\pi}{\ell_s^2} \int   dt dx^1 dx^2\, ,\\
\delta_\epsilon S_{\rm DBI}^{\rm D6} & = -\frac{8}{Rg_s} \frac{N_{\rm left} - N_{\rm right}}{|h|} {\cal V}_{\rm CY}\, e^{\frac{3z_0}{R}}   \frac{2\pi}{\ell_s^2} \int   dt dx^1 dx^2 =  \frac{8}{Rg_s} {\cal V}_{\rm CY}\, e^{\frac{3z_0}{R}}   \frac{2\pi}{\ell_s^2} \int   dt dx^1 dx^2\, .
\label{DBID6var}
\end{align}
That is, the dragging effect of the D6-branes ending on the D8-brane overcomes the effect of its tension, acting like an additional coupling $Q^{\rm eff}_{\rm D6} =  \frac{8}{3} e^{K/2} {\cal V}_{\rm CY}$.  This precisely compensates the coupling of the 4d membrane made up from a D8-brane in the case $\eta =1$ and from an anti-D8-brane in the case $\eta =-1$, as claimed. Microscopically, this cancellation is seen from the variation of the (anti-)D8-brane Chern-Simons action. By evaluating the coupling to the RR potential $C_9$ that corresponds to \eqref{610fluxes} and integrating over $X_6$ one obtains:
\be
S_{\rm CS}^{D8} = q_{\rm D8} \frac{2\pi}{\ell_s^9} \int C_9  =  -q_{\rm D8}\,  \eta \frac{5}{3} g_s^{-1}  {\cal V}_{\rm CY}  e^{\frac{3z_0}{R}} \, \frac{2\pi}{\ell_s^3} \int  dt dx^1 dx^2 \, .
\label{CSD8}
\ee
%with $q_{\rm D8} =1$ for a D8-brane and $q_{\rm D8} =-1$ for an anti-D8-brane. 
It is then easy to see that for $q_{\rm D8} = \eta$ the variation $\delta_\epsilon  S_{\rm CS}^{\rm D8}$ precisely cancels \eqref{DBID8var}+\eqref{DBID6var}. Therefore, the effect of the D6-branes can be understood as generating an effective coupling $Q_{\rm D8/D6}^{\rm eff} = Q_{\rm D8} + Q^{\rm eff}_{\rm D6} =  \eta q_{\rm D8}  e^{K/2} {\cal V}_{\rm CY}$. Indeed, notice that if one chose $q_{\rm D8} = - \eta$ then the Freed--Witten anomaly would be opposite and the D6-branes would be extending along $z \in (-\infty, z_0]$. This would result into $Q_{\rm D8/D6}^{\rm eff} = - e^{K/2} {\cal V}_{\rm CY}$, destabilising the system towards $z_0 \to -\infty$. 

\subsubsection*{Considering bound states}

In general, the Chern-Simons action of a D8-brane reads
\be
S_{\rm CS}^{\rm D8} =  \frac{2\pi}{\ell_s^9} \int P\left[ {\bf C} \wedge e^{-B}\right] \wedge e^{- \frac{\ell_s^2}{2\pi}  F} \wedge \sqrt{\hat{A}({\cal R})}\, ,
\ee
where ${\bf C} = C_1 + C_3 + C_5 + C_7 + C_9$ and $\hat{A}({\cal R}) = 1 + \frac{1}{24} \frac{\Tr R^2}{8\pi^2} + \dots $ is the A-roof genus. These couplings encode that in the presence of a worldvolume flux and/or curvature, we actually have a bound state of a D8 with lower-dimensional D-branes. If the bound state is BPS, then its tension will be a sum of D8 and D4-brane tensions. Taking also into account the effect of the D6-branes ending on it we have that
\be
T_{\rm D8}^{\rm total} = T_{\rm D8} + \left(K^F_a - K_a^{(2)}\right)  T_{\rm D4}^a\, ,
\label{TD8tot}
\ee
where $T_{\rm D4}^a = e^{K/2} t^a$, $K^F_a = \frac{1}{2\ell_s^6} \int_{X_6} \cF \wedge \cF \wedge \omega_a$ and $K_a^{(2)}$ has been defined in \eqref{Kcurv}. Similarly, the  Chern-Simons action of this bound state will give, upon dimensional reduction 
\be
q_{\rm D8} Q_{\rm D8}^{\rm total} = Q_{\rm D8/D6}^{\rm eff} + \left(K^F_a - K_a^{(2)}\right)  Q_{\rm D4}^a = \eta T_{\rm D8}^{\rm total}\, ,
\label{QD8tot}
\ee
where $ Q_{\rm D4}^a = \eta e^{K/2}   t^a$. Hence, again for $\eta=1$ a D8-brane will satisfy the BPS condition $Q=T$, while for $\eta =-1$ this will occur for an anti-D8-brane. One important aspect of these corrections is that the induced D4-brane tension in \eqref{TD8tot} is in general negative. Indeed, the curvature term $K_a^{(2)} t^a = -\frac{1}{24\ell_s^6} \int_{X_6} c_2(X_6) \wedge J_{\rm CY}$ is positive in the interior of the K\"ahler cone for Calabi--Yau geometries, inducing a negative D4-brane tension. In the present case this is compensated by an induced negative D4-brane charge in \eqref{QD8tot}. However, such a compensation will no longer occur for the non-supersymmetric AdS$_4$ flux backgrounds that we now turn to discuss.

%%%%%%%%%%%%%%%%%%%

\section{Non-supersymmetric AdS$_4$ vacua}
\label{s:nonsusy}

The type IIA flux potential obtained in  \cite{Grimm:2004ua} has, besides the supersymmetric vacua found in \cite{DeWolfe:2005uu}, further non-supersymmetric families of vacua. This can already be seen by the toroidal analysis of \cite{DeWolfe:2005uu,Camara:2005dc}, and it has been generalised to arbitrary Calabi--Yau geometries in \cite{Marchesano:2019hfb}. A subset of such vacua was analysed in \cite{Narayan:2010em} in terms of perturbative and non-perturbative stability, for the particular case of $X_6 = \mathbb{T}^6/(\Z_3 \times \Z_3)$. It was found that one particular family of vacua, dubbed type 2 in \cite{Narayan:2010em}, was stable both at the perturbative and non-perturbative level.\footnote{As pointed out in \cite{Marchesano:2019hfb} the remaining non-supersymmetric families (type 3 - type 8) found in \cite{Narayan:2010em} are not actual extrema of the flux potential, and only seem so when the potential is linearised as in \cite{Narayan:2010em}.} In the following we will extend this analysis to general Calabi--Yau geometries in the smearing approximation, and to new membranes like those arising from the D8/D6 configuration considered above. 

The non-supersymmetric vacua dubbed type 2 in \cite{Narayan:2010em} are in one-to-one correspondence with supersymmetric vacua, by a simple sign flip of the internal four-form flux $G_4 \to - G_4$. Because $G_4$ enters quadratically in the 10d supergravity Lagrangian, the energy of such a vacuum is similar to its  supersymmetric counterpart and, as argued in \cite{Narayan:2010em}, one expects it to share many of its nice properties. It was indeed found in \cite{Marchesano:2019hfb} that such non-supersymmetric vacua (dubbed {\bf A1-S1} therein) exist for any Calabi--Yau geometry, and that they are  stable at the perturbative level. Instead of the (smeared) supersymmetric relations \eqref{intfluxsm} we now have
\be
\ell_s [ H ]  = \frac{2}{5} m g_s [\re \Omega_{\rm CY} ] \, , \qquad  G_2  =  0\, ,  \qquad  \ell_s G_4  = \hat{e}_a  \tilde{\omega}^a  = \frac{3}{10} m \, \cK_a \tilde{\omega}^a  \, , \qquad  G_6 =  0\, , 
\label{intfluxsmnosusy}
\ee
and most features are analogous to the supersymmetric case. In particular, the AdS$_4$ radius and vacuum energy are also given by \eqref{Rads} and \eqref{lambda}, respectively. %Finally,  curvature corrections also redefine the flux quanta as in \eqref{curvflux}.

Because the energy dependence with the flux quanta is the same, one should be looking for similar non-perturbative transitions that jump to a vacuum of lower energy: Those that decrease $|\hat{e}_a|$ and those that increase $|m|$ or $|h|$. The objects that will implement such jumps will again be 4d membranes that come from (anti-)D4-branes and (anti-)D8-branes. Because of the sign flip in $G_4$, the role of the D4-branes will be exchanged with that of anti-D4-branes with respect to the supersymmetric case. 

Indeed, the relations \eqref{intfluxsmnosusy} imply that \eqref{610fluxes} is replaced by
\be
G_6 =   \frac{3\eta}{Rg_s} {\rm vol}_4 \wedge J_{\rm CY}\, , \qquad G_{10} =  - \frac{5\eta}{6 R g_s} {\rm vol}_4 \wedge J^3_{\rm CY}\, ,
\label{610fluxesnosusy}
\ee
with no further external fluxes. As a result we find the following 4d membrane couplings:
\be
Q_{\rm D2} = 0 \, , \qquad Q_{\rm D4}^{\rm ns} =  -  e^{K/2} \frac{\eta}{\ell_s^2} \int_\Sigma J_{\rm CY} \, , \qquad Q_{\rm D6} = 0\, , \qquad Q_{\rm D8} =  -\frac{5}{3}  \eta\, q_{\rm D8} e^{K/2} {\cal V}_{\rm CY}\, .
\label{QDGKTnosusy}
\ee
By analogy with the supersymmetric case, we now find that the equality $Q =T$ is realised by D4-branes wrapping anti-holomorphic two-cycles, for $\eta =1$, and holomorphic two-cycles for $\eta=-1$. This essentially amounts to exchanging the roles of D4-brane and anti-D4-brane, as advanced. If we chose the object with opposite charge (e.g. a D4-brane wrapping a holomorphic two-cycle for $\eta=1$) then we would have that $Q = -T$ and the effects of the tension and the coupling to the flux background would add up, driving the membrane away from the boundary. In general, it is not possible to find a D4-brane such that $Q >T$, just like it is not possible to find it in supersymmetric vacua. This reproduces the result of  \cite{Narayan:2010em} that D4-brane decays are, at best, marginal. Regarding D8-branes, the naive story is essentially the same as for ${\cal N} =1$ vacua. Since $Q_{\rm D8}$ remains the same, $Q^{\rm eff}_{\rm D8/D6}$ will compensate $T_{\rm D8}$ for $\eta=q_{\rm D8}$. 
%and the opposite charge of an anti-D8-brane for $\eta=-1$. 

Now, the interesting case occurs when we consider bound states of D8 and D4-branes, by introducing the effect of worldvolume fluxes and/or curvature corrections. In a D8-brane configuration similar to the one in the supersymmetric case the tension is the same:
\be
T_{\rm D8}^{\rm total} = T_{\rm D8} + \left(K^F_a - K_a^{(2)}\right)  T_{\rm D4}^a\, .
\label{TD8totnosusy}
\ee
In the large volume approximation $T_{\rm D8} \gg T_{\rm D4}^a$, and so just like in the supersymmetric case we need to consider a D8-brane whenever $\eta =1$, or else $T>Q$. The coupling of the corresponding 4d membranes is now different from \eqref{QD8tot}, and reads
\be
q_{\rm D8} Q_{\rm D8}^{\rm total} = Q_{\rm D8/D6}^{\rm eff} + \left(K^F_a - K_a^{(2)}\right)  Q_{\rm D4}^{{\rm ns}, a} = \eta \left[ T_{\rm D8} -  \left(K^F_a - K_a^{(2)}\right)  T_{\rm D4}^a\right]  \, .
\label{QD8totnosusy}
\ee
As a result we find that
\be
Q_{\rm D8}^{\rm total} - T_{\rm D8}^{\rm total} = 2 \left(K_a^{(2)} -K^F_a\right)  T_{\rm D4}^a\, ,
\label{QTsmnosusy}
\ee
where we have imposed $\eta=q_{\rm D8}$. On the one hand, by assumption the D8-brane worldvolume flux induces pure D4-brane charge, which means that $K^F_aT_{\rm D4}^a >0$.\footnote{If we consider  diluted fluxes that induce pure anti-D4-brane charge, their contributions would cancel in \eqref{QTsmnosusy}.} On the other hand,  generically $K_a^{(2)} T_{\rm D4}^a > 0$, since for a Calabi--Yau \cite{Miyaoka1987}
\be
-\int_{X_6} c_2(X_6) \wedge J_{\rm CY} \geq 0\, ,
\label{condcurv}
\ee
with the equality occurring only at the boundary of the K\"ahler cone. This means that the curvature corrections are inducing negative D4-brane charge and tension on the D8-brane. The effects of such negative tension and charge add up in the present non-supersymmetric background, and drag the D8-brane towards the AdS$_4$ boundary.\footnote{Notice that this mechanism is analogous to the one in  \cite{Maldacena:1998uz}, in which a D5-branes wraps the $K3$ in AdS$_3 \times S^3 \times K3$.} So if the worldvolume fluxes are absent or give a smaller contribution, we will have that $Q_{\rm D8}^{\rm total} > T_{\rm D8}^{\rm total}$ and the energy of the configuration will be minimised at $z_0 \to \infty$. As such, these D8/D4 bound states are clear candidates to realise the AdS instability conjecture of \cite{Ooguri:2016pdq,Freivogel:2016qwc}. In the next section we will argue that this is indeed the case.

While a remarkable result, one must realise that it does not apply to all non-supersymmetric vacua of this sort. It only applies to those flux vacua which contain space-time filling D6-branes, that is those with $N>0$ in \eqref{tadpole2}. If $N=0$ we cannot consider a transition like the above in which $m$ increases its absolute value. In other words, then the D8-brane configuration described above cannot exist.\footnote{Or it could at the expense of introducing anti-D6-branes, which would introduce a whole new set of instabilities.} These are precisely the kind of vacua considered in \cite{Narayan:2010em} which, even with these new considerations, would a priori remain marginally stable. Moreover, if \eqref{condcurv} vanished at some boundary of the K\"ahler cone, there would be a priori no instability triggered by D8/D4-brane bound states, which would be marginal.  In fact, this last statement is not true, but only a result of the smearing approximation. As we will see, when describing the same setup but in terms of a background that admits localised sources, corrections to the D8-brane tension will appear, modifying the above computation. 

%%%%%%%%%%%%%%%%%%%

\section{AdS$_4$ instability from the 4d perspective}
\label{s:insta}

The results of the previous section suggest that non-supersymmetric AdS$_4 \times X_6$ vacua with a flux background of the form \ref{intfluxsmnosusy} develop non-perturbative instabilities if they contains space-time filling D6-branes. From the 4d perspective such an instability would be mediated by a membrane that arises from wrapping a D8-brane on $X_6$, since it becomes a membrane with $Q > T$ upon dimensional reduction. However, the link between the inequality $Q > T$ and a non-perturbative gravitational instability typically follows an analysis similar to \cite{Maldacena:1998uz}, implicitly relying on the thin-wall approximation. As pointed out in \cite{Narayan:2010em}, D8-branes are not in the thin-wall approximation unless the value of $|m|$ is very large, which is not generically true. Therefore in this section we would like to provide an alternative argument of why these vacua are unstable. 

For this we will make use of the symmetry between supersymmetric and non-supersymmetric vacua mentioned in section \ref{s:nonsusy}. That is, for the same value of the fluxes $m$, $h$ and $|\hat{e}_a|$ the saxion vevs are stabilised at precisely the same value in both supersymmetric and non-supersymmetric vacua, and the vacuum energy \eqref{lambda} is also the same. For simplicity let us consider a pair of supersymmetric and non-supersymmetric vacua in which $e_0 = m^a=0$ and
\be
m^{\rm susy} = m^{\cancel{\rm susy}} > 0\, , \qquad  h^{\rm susy} = h^{\cancel{\rm susy}} > 0\, , \qquad \hat{e}_a^{\rm susy} = - \hat{e}_a^{\cancel{\rm susy}} > 0\, .
\ee
In both backgrounds, a D8-brane without worldvolume fluxes will induce the following shift of flux quanta as we cross it from $z = \infty$ to $z = -\infty$ as
\begin{align}
m^{\rm susy} \to m^{\rm susy} + 1\, , \qquad & |\hat{e}_a^{\rm susy}| \to |\hat{e}_a^{\rm susy}  + K_a^{(2)}| \, , \\
m^{\cancel{\rm susy}} \to m^{\cancel{\rm susy}} + 1\, , \qquad & |\hat{e}_a^{\cancel{\rm susy}}| \to  |\hat{e}_a^{\cancel{\rm susy}} + K_a^{(2)}| = |\hat{e}_a^{{\rm susy}}  - K_a^{(2)}|\, .
\end{align}
Because the absolute value of the four-form flux quanta $\hat{e}_a$ are different after the jump for the supersymmetric and the non-supersymmetric case, so are the vevs of the K\"ahler moduli and the vacuum energy. To fix this, let us add to the supersymmetric setup a D4-brane wrapping a holomorphic two-cycle in the Poincar\'e dual class to $2K_a^{(2)}[\tilde{\omega}^a]$. The resulting 4d membrane can create a marginal bound state with the one coming from the D8-brane, implementing the combined jump
\begin{align}
m^{\rm susy} \to m^{\rm susy} + 1\, , \qquad  |\hat{e}_a^{\rm susy}| \to |\hat{e}_a^{\rm susy}  - K_a^{(2)}| \, .
\end{align}
Now both supersymmetric and non-supersymmetric jumps are identical, in the sense that the variation of the scalar fields from the initial to the final vacuum is the same, and so are the initial and final vacuum energies. As a result, the energy stored in the field variation of both solutions should be identical. What is different is the tension of the membranes. We have that
\be
T_{\rm susy} = T_{\rm D8} +  K_a^{(2)}  T_{\rm D4}^a > T_{\rm D8} -  K_a^{(2)}  T_{\rm D4}^a = T_{\cancel{\rm susy}}\, ,
\ee
assuming as before that \eqref{condcurv} is met. Therefore, because the supersymmetric decay is marginal, the non-supersymmetric one should be favoured energetically, rendering the non-supersymmetric vacuum unstable. 

%%%%%%%%%%%%%%%%%%%

\section{Beyond the smearing approximation}
\label{s:nonsmeared}

The Calabi--Yau flux backgrounds of section \ref{s:dgkt} and \ref{s:nonsusy} can be thought of as an approximation to the actual 10d solutions to the equations of motion and Bianchi identities, in which O6-planes and D6-branes are treated as localised sources. More precisely, the smeared Calabi--Yau solution can be recovered from the actual solution in the limit of small cosmological constant, weak string coupling and large internal volume \cite{Saracco:2012wc,Junghans:2020acz,Marchesano:2020qvg}. Any of these quantities can be used to define an expansion parameter, so that the actual 10d solution can be described as a perturbative series, of which the smeared solution is the leading term. While a solution for the whole series (i.e. the actual 10d background) has not been found yet, the next-to-leading term of the expansion was found in \cite{Marchesano:2020qvg} for the case of the supersymmetric vacua. In the following we will review the main results of \cite{Marchesano:2020qvg}, and then use the approach of \cite{Junghans:2020acz} to construct, at the same level of accuracy, a similar background with localised sources for the non-supersymmetric vacua of section \ref{s:nonsusy}. As we will see, these more precise backgrounds do not affect significantly the energetics of 4d membranes made up from D4-branes. However, as it will be discussed in the next section, they yield non-trivial effects for membranes that correspond to D8/D6 systems.

\subsection{Supersymmetric AdS$_4$}
\label{ss:mpqt}

To incorporate localised sources to the type IIA flux compactification of section \ref{ss:smeared} one must first consider a warped metric of the form
\begin{equation}\label{eq:warped-product}
	ds^2 = e^{2A}ds^2_{\mathrm{AdS}_4} + ds^2_{X_6}\, ,
\end{equation}
 with $A$ a function on $X_6$. Then, as pointed out in \cite{Marchesano:2020qvg}, the Calabi--Yau metric on $X_6$ must be deformed to a metric that corresponds to an $SU(3) \times SU(3)$-structure solution with $G_6 =0$. Assuming as before that P.D.$[\ell_s^{-2}H] = h [\Pi_{\rm O6}] = h [\Pi_{\rm D6}]$,  the first-order correction to the smearing approximation can be described in terms of the following equation
\be
\ell_s^2 \Delta_{\rm CY} K =  \frac{2}{5} m^2 g_s \re \Omega_{\rm CY} + (N-4)\d(\Pi_{\rm O6})\, ,
\label{defK}
\ee
where $\Delta_{\rm CY} = d^\dag_{\rm CY} d + d d^\dag_{\rm CY}$ is constructed from the CY metric. The solution is of the form
\be
 K = \varphi \re \Omega_{\rm CY}  + \re k \, ,
\label{formK}
\ee
with $k$ a (2,1) primitive current and $\varphi$ is a real function that satisfies $\int_{X_6} \varphi = 0$ and
\be
\Delta_{\rm CY}  \varphi = \frac{mh}{4}\left(\frac{{\cal V}_{\Pi_{\rm O6}}}{{\cal V}_{\rm CY}} - \delta^{(3)}_{\Pi_{\rm O6}}\right)  \ \implies \ \varphi \sim \cO(g_s^{1/3})\, ,
\ee
where $\delta^{(3)}_{\Pi_{\rm O6}} \equiv *_{\rm CY} (\im \Omega_{\rm CY} \wedge \d(\Pi_{\rm O6}))$. In term of these quantities we can describe the metric background and the varying dilaton profile as
\begin{subequations}	
	\label{solutionsu3}
\begin{align}
J & = J_{\rm CY} + \cO(g_s^2) \, , \qquad   \qquad  \Omega  = \Omega_{\rm CY} + g_s k +  \cO(g_s^2)\, , \\
e^{-A}  & = 1 + g_s \varphi + \cO(g_s^2) \, , \qquad e^{\phi}   = g_s \left(1 - 3  g_s \varphi\right) + \cO(g_s^3)\, .
\end{align}
\end{subequations}   
where we have taken $g_s$ as the natural expansion parameter. Notice that $\varphi \sim -\frac{mh}{4r}$ near $\Pi_{\rm O6}$, and so as expected the 10d string coupling blows up and the warp factor becomes negative near that location. The function $\varphi$ indicates the region $\tilde{X}_6 \equiv \{p \in X_6 | g_s |\varphi(p)| \ll 1\}$ in which the perturbative expansion on $g_s$ is reliable; beyond that point one may use the techniques of \cite{DeLuca:2021mcj} to solve  the 10d supersymmetry equations. The background fluxes are similarly expanded as
\begin{subequations}
	\label{solutionflux}
\begin{align}
 H & =   \frac{2}{5} \frac{m}{\ell_s} g_s \left(\re \Omega_{\rm CY} + g_s K \right) - \oh   d\re \left(\bar{v} \cdot \Omega_{\rm CY} \right) + \cO(g_s^{3}) \label{H3sol} \, , \\
 \label{G2sol}
 G_2 & =     d^{\dag}_{\rm CY} K  + \cO(g_s)  = - J_{\rm CY} \cdot d(4 \varphi \im \Omega_{\rm CY} - \star_{\rm CY} K) + \cO(g_s) \, , \\
G_4 & =  \frac{m}{\ell_s} J_{\rm CY} \wedge J_{\rm CY} \left(\frac{3}{10}  - \frac{4}{5} g_s \varphi \right)+   J_{\rm CY} \wedge g_s^{-1} d \im v + \cO(g_s^2) \, , \\
G_6 & = 0\, .
\end{align}
\end{subequations}   
Here $v$ is a (1,0)-form whose presence indicates that we are in a genuine $SU(3)\times SU(3)$ structure, as opposed to an $SU(3)$ structure. It is determined by
\be
v  = g_s \p_{\rm CY} f_\star + \cO(g_s^3)\, , \qquad \text{with} \qquad \ell_s \Delta_{\rm CY} f_\star  = - g_s 8 m \varphi \, .
\ee 
It is easy to see that \eqref{solutionsu3} and \eqref{solutionflux} reduce to the smeared solution in the limit $g_s \to 0$.  Moreover, as shown in \cite{Marchesano:2020qvg}, this background satisfies the supersymmetry equations and the Bianchi identities up to order $\cO(g_s^2)$.  As a cross-check of this result, we discuss in Appendix \ref{ap:10deom} how the 10d equations of motion are satisfied,  to the same level of accuracy.

Given this new background, one may reconsider the computation of the tension and coupling made in the smearing approximation. Let us for instance consider a D4-brane wrapping a two-cycle $\Sigma$. Instead of the expression for $G_6$ in \eqref{610fluxes} we obtain
\bea\nonumber
G_6 & = & -  {\rm vol}_4 \wedge \left[ J_{\rm CY}  \frac{m}{5\ell_s } \left(3 - 8 g_s \varphi \right) - \oh *_{\rm CY} d \left(J_{\rm CY} \wedge d^c f_\star \right)  \right] e^{4A}+ \cO(g_s^2) \\ \nonumber
& =& -  {\rm vol}_4 \wedge \left[ J_{\rm CY}  \frac{m}{5\ell_s } \left(3 - 20 g_s \varphi \right) - \oh \left( \Delta_{\rm CY} - dd^\dag_{\rm CY}\right) \left(f_\star  J_{\rm CY}  \right)  \right] + \cO(g_s^2) \\
& =&  -  {\rm vol}_4 \wedge \left[ J_{\rm CY} \frac{3\eta}{Rg_s} + \oh dd^\dag_{\rm CY} \left(f_\star  J_{\rm CY}  \right)  \right] + \cO(g_s^2)\, ,
\label{6fluxesloc}
\eea
where $d^c \equiv i(\bar{\p}_{\rm CY} - \p_{\rm CY})$ and we have used that $J_{\rm CY} \wedge d^c f  = *_{\rm CY} d(J_{\rm CY} f)$. Since the only difference with respect to the smearing approximation is an exact contribution, the membrane coupling $Q_{\rm D4}$ remains unchanged, and it is still given by $Q_{\rm D4} =  \eta e^{K/2} \int_\Sigma J_{\rm CY}$. As before, D4-branes  wrapping holomorphic ($\eta=1$) and anti-holomorphic ($\eta=-1$) two-cycles will be BPS, and will feel no force in the above AdS$_4$ background, as expected from supersymmetry.

\subsection{Non-supersymmetric AdS$_4$}
\label{ss:nonsmearednonsusy}

Just like for supersymmetric vacua, one would expect a 10d description of the non-supersymmetric vacua of section \ref{s:nonsusy} compatible with localised sources. Again, the idea would be that the smeared background is the leading term of an expansion in powers of $g_s$. In the following we will construct a 10d background with localised sources which can be understood as a first-order correction to the smeared Calabi--Yau solution \eqref{intfluxsmnosusy} in the said expansion. 

The main feature of the non-supersymmetric background \eqref{intfluxsmnosusy} is that it flips the sign of the RR four-form flux $G_4$, while it leaves the remaining fluxes, metric, dilaton and vacuum energy invariant. This means that the Bianchi identities \eqref{BIG} do not change at leading order, and in particular the leading term for two-form flux $G_2$ should have the same form \eqref{G2sol} as in the supersymmetric case. Moreover, the localised solution is likely to be described in terms of the quantities $\varphi$ and $k$ that arise from the Bianchi identity of $G_2$, at least at the level of approximation that we are seeking. Because of this, it is sensible to consider a 10d metric and dilaton background similar to the supersymmetric case, namely \eqref{solutionsu3}. 

Regarding the background flux $G_4$, there should be a sign flip on its leading term, but it is clear that this cannot be promoted to an overall sign flip, because the co-exact piece of $G_4$, that contributes to the Bianchi identity, must be as in the supersymmetric case. Since the harmonic and co-exact pieces of the fluxes are fixed by the smearing approximation and the Bianchi identities, the question is then how to adjust their exact pieces to satisfy the equations of motion. Using the approach of \cite{Junghans:2020acz}, we find that the appropriate background reads
\begin{subequations}
	\label{solutionfluxnnosusy}
\begin{align}
 H & =   \frac{2}{5} \frac{m}{\ell_s} g_s \left(\re \Omega_{\rm CY} - 2g_s K \right) + \frac{1}{10}   d\re \left(\bar{v} \cdot \Omega_{\rm CY} \right) + \cO(g_s^{3})  \, , \\
 G_2 & =     d^{\dag}_{\rm CY} K  + \cO(g_s)   \, , \\
G_4 & =  -\frac{m}{\ell_s} J_{\rm CY} \wedge J_{\rm CY} \left(\frac{3}{10}  + \frac{4}{5} g_s \varphi \right) -\frac{1}{5}   J_{\rm CY} \wedge g_s^{-1} d \im v + \cO(g_s^2) \, , \\
G_6 & = 0\, ,
\end{align}
\end{subequations} 
with the same definition for the (1,0)-from $v$. In Appendix \ref{ap:10deom} we show that this background satisfies the 10d equations of motion up to order $\cO(g_s^2)$, just like the supersymmetric case.

With this solution in hand, one may proceed as in the supersymmetric case and recompute the 4d membrane couplings and tensions. If the result is different from the one in the smearing approximation the difference could be interpreted as a $g_s$ correction. To begin, let us again consider a D4-brane wrapping a two-cycle $\Sigma$. The coupling of such a brane can be read from the six-form RR flux with legs along AdS$_4$, which reads
\bea\nonumber
G_6 & = & {\rm vol}_4 \wedge \left[ J_{\rm CY}  \frac{m}{5\ell_s } \left(3 + 8 g_s \varphi \right) + \frac{1}{10} *_{\rm CY} d \left(J_{\rm CY} \wedge d^c f_\star \right)  \right] e^{4A}+ \cO(g_s^2) \\ \nonumber
& =& {\rm vol}_4 \wedge \left[ J_{\rm CY}  \frac{m}{5\ell_s } \left(3 - 4 g_s \varphi \right) + \frac{1}{10} \left( \Delta_{\rm CY} - dd^\dag_{\rm CY}\right) \left(f_\star  J_{\rm CY}  \right)  \right] + \cO(g_s^2) \\
& =&  {\rm vol}_4 \wedge \left[ J_{\rm CY} \frac{3\eta}{Rg_s} - \frac{1}{10} dd^\dag_{\rm CY} \left(f_\star  J_{\rm CY}  \right)  \right] + \cO(g_s^2)\, .
\label{6fluxeslocnosusy}
\eea
Remarkably, we again find that the first non-trivial correction to the smearing approximation is an exact form, and so it vanishes when integrating over $\Sigma$. As a result, the 4d membrane couplings $Q_{\rm D4}^{\rm ns} =  - \eta e^{K/2} \ell_s^{-2}\int_\Sigma J$ remain uncorrected at this level of the expansion, and there is a force cancellation for D4-branes wrapping anti-holomorphic ($\eta=1$) and holomorphic ($\eta=-1$) two-cycles, just like in our discussion of section  \ref{s:nonsusy}. Presumably, by looking at higher-order corrections one may find one that violates the equality $Q_{\rm D4}^{\rm ns} = T_{\rm D4}^{\rm ns}$ in one way or the other, which would be a non-trivial test of the conjecture in \cite{Ooguri:2016pdq}. Such a computation is however beyond the scope of the present work. Instead, we will focus on membranes whose coupling and tension departure from the smeared result already at this level of approximation, namely those membranes arising from D8/D6 systems. To see how this happens, one must first take into account that beyond the smearing approximation such systems are described by BIonic configurations, as we now discuss. 

%%%%%%%%%%%%%%%%%%%

\section{BIonic membranes}
\label{s:bion}

A D$p$-brane ending on a D$(p+2)$-brane to cure a Freed--Witten anomaly constitutes a localised source for gauge theory on the latter. When going beyond the smearing approximation one should take this into account, and describe the combined system as a BIon-like solution \cite{Gibbons:1997xz}. In this section we do so for the D8/D6-brane system, and compute the  tension and flux coupling of the associated 4d membrane for both the supersymmetric and non-supersymmetric backgrounds of the last section. As we will see, the BIonic nature of the membrane will modify their coupling and tension of the membrane with respect to the smeared result.

\subsection{Supersymmetric AdS$_4$}
\label{ss:bionsusy}

Let us consider a D8-brane wrapping $X_6$ with orientation $q_{\rm D8} = \pm 1$ and extended along the plane $z = z_0$ in the Poincar\'e patch of AdS$_4$. As pointed out above, due to the non-trivial $H$-flux background we must have an excess of $h$ D6-branes wrapping $\Pi_{\rm O6}$ and extended to the right of the D8-brane, namely along $(t,x^1,x^2) \times [z_0 , \infty) \subset$ AdS$_4$. This setup implies a Bianchi identity for the D8-brane worldvolume flux $\cF = B + \frac{\ell_s^2}{2\pi} F$ of the form
\be
d{\cal F} = H - \frac{h}{\ell_s} \delta(\Pi_{\rm O6})\, .
\label{dFD8}
\ee
Because by construction the rhs is trivial in cohomology, this equation always has a solution. Moreover, if we are in the smearing approximation, we have that the rhs of \eqref{dFD8} vanishes, and so $\cF$ must be closed. The energy-minimising configurations then correspond to solving the standard F-term and D-term-like equations for $\cF$ \cite{Koerber:2007jb}, which in our setup means that $\cF$ is a harmonic (1,1)-form of $X_6$ such that $3\cF \wedge J_{\rm CY}^2  = \cF^3$. When we see such a D8-brane as a membrane in 4d, this harmonic worldvolume flux is the one responsible for the contribution $K_a^F Q^a_{\rm D4}$ to their flux coupling and tension. 

If we describe our system beyond the smearing approximation, the D8-brane worldvolume flux can no longer be closed. Instead, it must satisfy a Bianchi identity that is almost identical to the one of the RR two-form flux. Even when the harmonic piece of $\cF$ vanishes, we find that
\be
\cF =  \frac{G_2}{G_0} = \frac{\ell_s}{m} d^{\dag}_{\rm CY} K  + \cO(g_s) \, .
\label{cfsol}
\ee
assuming that the D6-branes are equally distributed on top of the O6-planes before and after the jump, see \cite{Casas:2022mnz} for more general setups.  BPS configurations with D$p$-branes ending on D$(p+2)$-branes, inducing a non-closed worldvolume flux on the latter are usually described by BIon-like solutions \cite{Gibbons:1997xz}, in which the D$(p+2)$-brane develops a spike along the direction in which the D$p$-branes are extended. A relatively simple configuration of this sort is given by the D5/D3 system in type IIB flux compactifications, that was analysed in \cite{Evslin:2007ti} from the viewpoint of calibrations. In this setup a D5-brane wraps a special Lagrangian three-cycle $\Lambda$ of a warped Calabi--Yau compactification, and extends along the plane $x^3 = x^3_0$ of $\pr^{1,3}$. If $\int_{\Lambda} H = - N$, then $N$ space-time filling D3-branes must end on the D5-brane, stretched along $(t,x^1,x^2) \times [x^3_0 , \infty) \subset \pr^{1,3}$ and located at a point $p \in \Lambda$. This induces an internal worldvolume flux on the D5-brane, solving the equation $d\cF = N\left(\delta(p) - \frac{d{\rm vol}_\Lambda}{{\rm Vol}(\Lambda)}\right)$. To render the configuration BPS it is necessary to give a non-trivial profile to the D5-brane position field $X^3$, such that $dX^3 = *_{\Lambda} \cF$. The resulting profile features a spike $X^3 \sim \frac{N}{r}$ around the point $p$, which represents the $N$ D3-branes ending on the D5. The D5-brane BIon configuration accounts for the whole energy of the D5/D3 system. 

\begin{figure}[h!]
    \centering
    \includegraphics[width=10cm]{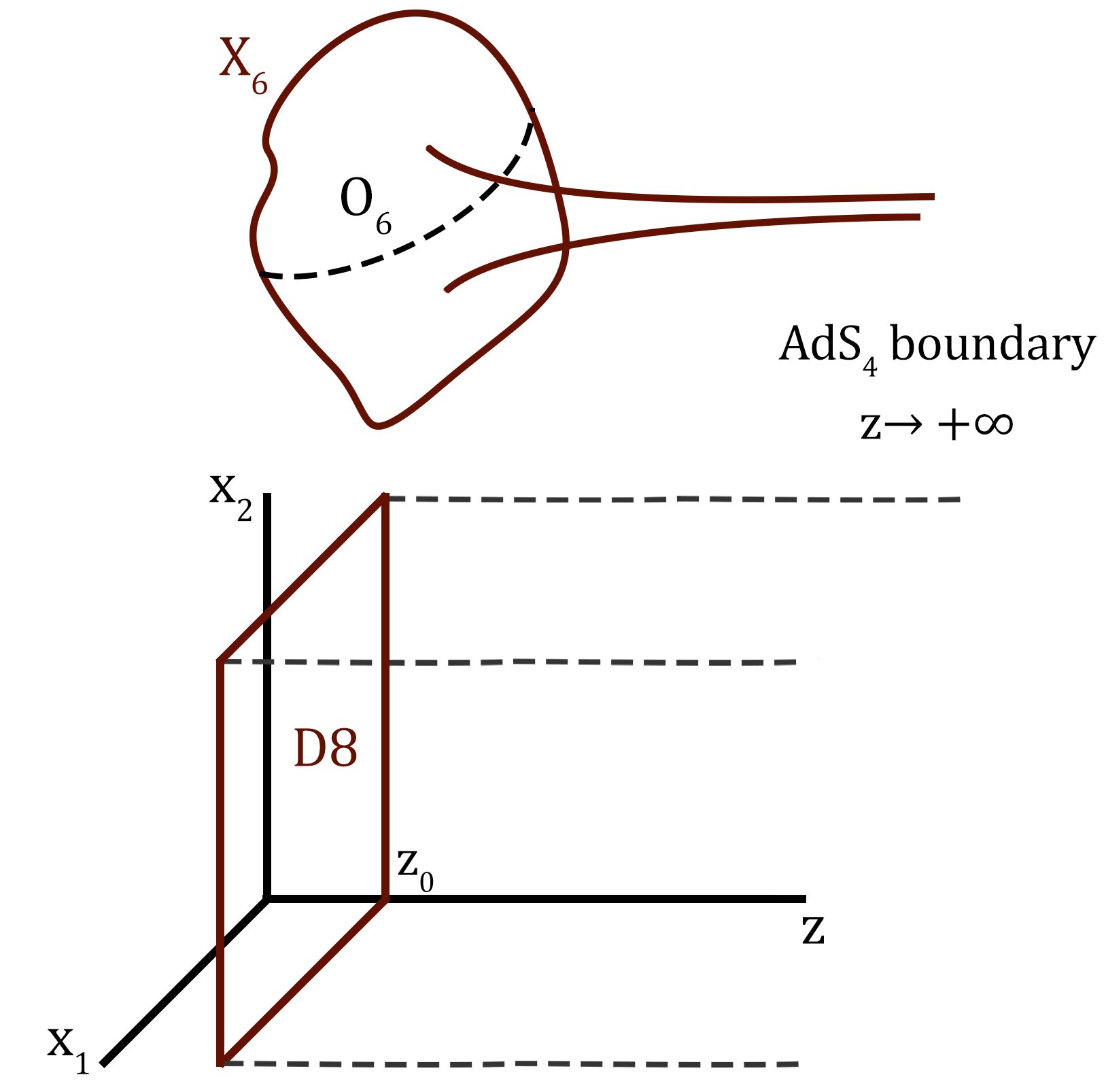}
    \caption{Beyond the smearing approximation, the D8/D6 system of figure \ref{fig:D8D6} becomes a BIon-like solution for the D8-brane, with a BIon profile that peaks at the O6-plane location.}
    \label{fig:D8D6bion}
\end{figure}

Our D8/D6 setup can be seen as a six-dimensional analogue of the D5/D3 system. The presence of the worldvolume flux \eqref{dFD8} can be made compatible with a BPS configuration if we add a non-trivial profile for the D8-brane transverse field $Z$. The relation with the worldvolume flux is now given by
\be
*_{\rm CY} dZ = q_{\rm D8} \im \Omega_{\rm CY} \wedge \cF + \cO(g_s) \, .
\label{BIonrel}
\ee
This expression can be motivated in a number of ways. In Appendix \ref{ap:dbi} we show that upon imposing it the DBI action is linearised at the level of approximation that we are working, as required by a BPS configuration. In Appendix \ref{ap:IIBion} we describe a similar configuration in type IIB flux compactifications, that can then be mapped to the BPS Abelian SU(4) instantons of Calabi--Yau four-folds \cite{Donaldson:1996kp}. Finally, notice that \eqref{BIonrel} implies that
\be
\Delta_{\rm CY} Z = \ell_s q_{\rm D8} h \left( \delta^{(3)}_{\Pi_{\rm O6}} -\frac{{\cal V}_{\Pi_{\rm O6}}}{{\cal V}_{\rm CY}} \right)\, ,
\ee
and so whenever $q_{\rm D8} h = |h|$ we recover a spike profile of the form $Z \sim \frac{|h|\ell_s}{r}$ near $\Pi_{\rm O6}$, as expected. In fact we can draw the more precise identification
\be
Z = z_0 - \frac{4\ell_s \varphi}{|m|}\, ,
\ee
where we have imposed the BPS relation $q_{\rm D8} = \eta \equiv {\rm sign}\, m$. Notice that this identifies the spike profile of the BIon solution towards the AdS$_4$ boundary with the strong coupling region near the O6-plane location, where our perturbative  expansion on $g_s$ is no longer trustable, see fig. \ref{fig:D8D6bion}.

The relation \eqref{BIonrel} implies that the DBI action of the BIon can be computed in terms of calibrations. Indeed, ignoring  curvature corrections, the calibration for a D8-brane wrapping $X_6$ and with worldvolume fluxes is given by 
\be
- \im \Phi_+ = - g_s^{-1} q_{\rm D8} \im e^{-iJ_{\rm CY}} + \cO(g_s) = g_s^{-1} q_{\rm D8}\left(- \frac{1}{6} J_{\rm CY}^3 + J_{\rm CY}\right) + \cO(g_s)\, ,
\label{imPhi+}
\ee
while that for D6-branes wrapping a three-cycle of $X_6$ is 
\be
\im \Phi_- =  g_s^{-1} \left( \im v + \Im \Omega - \oh \psi \im \omega \wedge \im \Omega  \right) + \cO(g_s)  = g_s^{-1} \Im \Omega_{\rm CY} + \cO(g_s^0)\, ,
\label{imPhi-}
\ee
at leading order in our expansion.  Here $\psi$ and $\omega_0$ are a complex function and 2-form which describe the $SU(3)\times SU(3)$ structure, and such that $\Omega = \frac{i}{\psi} v \wedge \omega + \cO(g_s^2)$, see \cite{Marchesano:2020qvg} for details. Applying the general formulas of \cite{Evslin:2007ti}, we find that the BIon DBI action reads %for $q_{\rm D8} =1$ reads\footnote{For $q_{\rm D8} = -1$ we get an overall minus sign, which is cancelled by integrating over the opposite orientation of $X_6$. }
\bea
\label{psp}
dS_\text{DBI}^{\rm D8} &= & dt \wedge dx^1 \wedge dx^2 \wedge e^{\frac{3Z}{R}} q_{\rm D8} \left(  \im \Phi_+ - dZ \wedge e^A \im \Phi_-  \right) \wedge e^{-\cF} \\
& \simeq & dt \wedge dx^1 \wedge dx^2 \wedge g_s^{-1} e^{\frac{3z_0}{R}}  \left( \frac{1}{6}J_{\rm CY}^3  - \oh J_{\rm CY} \wedge \cF^2 + q_{\rm D8}dZ \wedge \im \Omega_{\rm CY} \wedge \cF \right)  \\
& = &   - dt \wedge dx^1 \wedge dx^2 \wedge g_s^{-1} e^{\frac{3z_0}{R}}  \left(- \frac{1}{6}J_{\rm CY}^3  + \oh J_{\rm CY} \wedge \cF^2 + *_{\rm CY} dZ \wedge dZ \right)\, .
\label{finaldbi}
\eea
The last line coincides with our result of Appendix \ref{ap:dbi}, and with what is expected for a BIon solution. Indeed, the first two terms of \eqref{finaldbi} correspond to the DBI action of the magnetised D8-brane, while the third one corresponds to the D6-branes that stretch towards the AdS$_4$ boundary. Nevertheless, notice that the middle term $\oh J_{\rm CY} \wedge \cF^2$ gives an extra contribution to the DBI action compared to the smearing approximation of section \ref{ss:4dmem}. Indeed, when $\cF$ is a harmonic form this term accounts for the contribution $K_a^FT_{\rm D4}^a$ in \eqref{TD8tot}. When going away from the smearing approximation $\cF$ will also have a co-exact piece, given by \eqref{cfsol}, that will contribute to the DBI even if $\cF^{\rm harm} =0$. Because it induces a non-trivial D4-brane charge, one may interpret this extra contribution to the D8-brane tension as a curvature correction induced by the non-trivial BIon profile, as opposed to D6-branes sharply ending on the D8-brane, although it would be interesting to derive this expectation from first principles. As we will see, this additional contribution to the tension does not play much of a role in the present supersymmetric setup, but it is crucial for the dynamics of Bionic membranes in non-supersymmetric backgrounds. 

Eq.\eqref{psp} suggests how to generalise \eqref{BIonrel} to a relation describing the BIon profile to all orders in $g_s$. The natural choice is 
\be
  *_6 dZ =  -  q_{\rm D8} e^{\phi-2A} \left. \im \Phi_- \wedge  e^{-\cF}\right|_5 \, ,
  \label{stardZ}
\ee
where the Hodge star is performed with the exact, non-Calabi--Yau metric of $X_6$, and $|_5$ means that we are only keeping the five-form component of the polyform on the rhs. With this choice the BIon DBI action would read
\be
dS_\text{DBI}^{\rm D8} =
 dt \wedge dx^1 \wedge dx^2 \wedge  e^{\frac{3Z}{R}} q_{\rm D8} \left(  \im \Phi_+  \wedge e^{-\cF} - e^{3A-\phi} *_{6} dZ \wedge dZ \right)\, ,
\label{finaldbiex}
\ee
as expected on general grounds. In addition, \eqref{psp} encodes the force cancellation observed for the D8/D6 system in the smearing approximation, which can now be derived for the single object which is the BIonic D8-brane, and in the exact background. For this, notice that the Chern-Simons part of the D8-brane action reads
\be  
dS_\text{CS}^{\rm D8}  = - dt \wedge   dx^1 \wedge dx^2 \wedge \frac{R}{3} e^{\frac{3Z}{R}} e^{4A} q_{\rm D8}  *_6 \lambda \hat{G}  \wedge e^{-\cF}  \, ,
\ee
where $\hat{G}$ is defined as in \eqref{demoflux}. Putting both contributions together and using the bulk supersymmetry equation
\be
 d_H  \left( e^A \im \Phi_- \right) +  \frac{3}{R} \im \Phi_+ =  e^{4A} *_6 \lambda \hat{G}\, ,
 \label{susy1}
\ee
and \eqref{dFD8} one finds that 
\begin{align}
\label{DBICS}
  dS_\text{DBI}^{\rm D8}  & + dS_\text{CS}^{\rm D8} = - dt \wedge dx^1 \wedge dx^2 \wedge \frac{R}{3} q_{\rm D8}  \left[ de^{\frac{3Z}{R}} \wedge  e^A \im \Phi_- +  e^{\frac{3Z}{R}}  d_H   \left( e^A \im \Phi_- \right)  \right]  \wedge e^{-\cF} \\ \nonumber
& =   - dt \wedge dx^1 \wedge dx^2 \wedge \frac{R}{3} q_{\rm D8}  \left[ d \left(e^{\frac{3Z}{R}}   e^A \im \Phi_- \wedge e^{-\cF} \right) + \frac{h}{\ell_s} e^{\frac{3Z}{R}}  \delta(\Pi_{\rm O6})  \wedge e^A \im \Phi_- \wedge  e^{-\cF} \right]  \, .
\end{align}
The first term of the second line is a total derivative that will vanish when integrating over $X_6$, while the second term is an infinite contribution to the action, that accounts for the DBI action of the $|h|$ D6-branes extending along $[z_0, \infty)$. Indeed, it is easy to see that the leading piece of this term is of the form $ |h| g_s^{-1} e^{\frac{3Z}{R}} \delta(\Pi_{\rm O6})  \wedge \im \Omega_{\rm CY} = |h|g_s^{-1} {\cal V}_{\Pi_{\rm O6}} e^{\frac{3Z_\infty}{R}}$, with $Z_\infty \equiv Z|_{\Pi_{\rm O6}} = \infty$. The relevant point is that $Z_{\infty}$ is independent of $z_0$, and therefore this second term is independent of the D8-brane transverse position. Therefore, the total energy of the BIonic 4d membrane will be independent of $z_0$, even if contains some infinite contributions. This matches the results obtained in the smearing approximation, and is equivalent to the BPS equilibrium relation $Q_{\rm D8}^{\rm BIon} = T_{\rm D8}^{\rm BIon}$. %Finally, it is easy to see that the computation is similar for the case of $\eta = q_{\rm D8} = -1$, with an overall sign change in \eqref{DBICS}.

The above computation is quite general, and essentially follows from some  general observations made in \cite{Koerber:2007jb} applied to the present setup. It is nevertheless instructive to see how \eqref{susy1}, which is a key relation to achieve force cancellation for our BIonic D8-brane, is satisfied for the background \eqref{solutionsu3} and \eqref{solutionflux}, in preparation for the non-supersymmetric case. We have that
\begin{align}
&d_H  \left( e^A \im \Phi_- \right)  = \frac{1}{2}d d^c f_\star + *_{\rm CY} G_2 - \frac{2}{3} G_0 \left(\frac{2}{5}- g_s\varphi \right)J_{\rm CY}^3  +  \cO(g_s^{5/3})\, , \\
 &\frac{3}{R} \im \Phi_+  = \frac{3}{5} q_{\rm D8}  |G_0| \left( - J_{\rm CY} + \frac{1}{6} J_{\rm CY}^3\right) + \cO(g_s^{2})\, , \\ 
 & e^{4A} *_6 \lambda \hat{G}  = -\frac{1}{2} dd^\dag_{\rm CY} \left(f_\star  J_{\rm CY}  \right) -\frac{3}{5} G_0 J_{\rm CY} - *_{\rm CY} G_2 - \frac{1}{6} G_0 \left(1 - 4g_s \varphi\right) J_{\rm CY}^3 + \cO(g_s^{5/3})\, , 
\end{align}
and so one only has to impose $\eta=q_{\rm D8} $ and use that $d d^c f = - dd^\dag_{\rm CY} \left(f  J_{\rm CY}  \right)$  to show the equality. 

\subsection{Non-supersymmetric AdS$_4$}
\label{ss:bionnosusy}

Let us now consider the D8-brane BIon in the non-supersymmetric AdS$_4$ background of section \ref{ss:nonsmearednonsusy}. Notice that the metric and dilaton background are similar to  the supersymmetric case, and that the $H$-flux only changes by an exact piece at subleading order, so that \eqref{cfsol} remains intact. Because of this, the DBI action of the BIon should be identical to the supersymmetric case, at least to the level of approximation that we are working, and so should be the BIon profile \eqref{BIonrel}. One may thus run a very similar argument to \eqref{DBICS} to see whether the D8-brane is in equilibrium or not with the background. If not, the same computation will determine whether it is dragged towards the boundary or away from it. 

The key relation to look at is again the bulk supersymmetry equation \eqref{susy1}. If  satisfied, the BIonic membrane will be at equilibrium for any choice of transverse position $z_0$. In the smearing approximation we have already seen that there is no equilibrium whenever there is a non-trivial D4-brane charge induced by curvature or worldvolume fluxes, c.f. \eqref{QTsmnosusy}, so we do not expect \eqref{susy1} to be satisfied. Evaluating the background \eqref{solutionsu3} and \eqref{solutionfluxnnosusy} one finds that
\begin{align}
&d_H  \left( e^A \im \Phi_- \right)  = \frac{1}{2}d d^c f_\star + *_{\rm CY} G_2 - \frac{2}{15} G_0 \left(2 +  g_s\varphi \right)J_{\rm CY}^3  +  \cO(g_s^{4/3})\, , \\
 &\frac{3}{R} \im \Phi_+  = \frac{3}{5} q_{\rm D8}  |G_0| \left( - J_{\rm CY} + \frac{1}{6} J_{\rm CY}^3\right) + \cO(g_s^{2})\, , \\ 
 & e^{4A} *_6 \lambda \hat{G}  = \frac{1}{10} dd^\dag_{\rm CY} \left(f_\star  J_{\rm CY}  \right) + \frac{3}{5} G_0 J_{\rm CY} - *_{\rm CY} G_2 - \frac{1}{6} G_0 \left(1 - 4g_s \varphi\right) J_{\rm CY}^3 + \cO(g_s^{5/3})\, , 
\end{align}
which results in\footnote{In the language of \cite{Lust:2008zd,Held:2010az}, this corresponds to a background where gauge BPSness is not satisfied, and as a result some space-time filling D-branes may develop tachyons. One can however check that D6-branes wrapping special Lagrangians of $X_6$, and in particular those on top of the orientifold, do not develop any instability. It would be interesting to see if D8-branes wrapping coisotropic five-cycles \cite{Font:2006na} could develop them.}
\be
 d_H  \left( e^A \im \Phi_- \right) +  \frac{3}{R} \im \Phi_+ -  e^{4A} *_6 \lambda \hat{G} = - \frac{3}{5} dd^\dag \left(f_\star  J_{\rm CY}  \right) -\frac{6}{5} G_0 J_{\rm CY}  - \frac{4}{5}  G_0 g_s \varphi J_{\rm CY}^3  +\cO(g_s^{4/3})\, .
\ee
Plugged into the DBI and CS actions, and again ignoring curvature terms, this translates into
\begin{align}
\nonumber
  dS_\text{DBI}^{\rm D8}   + dS_\text{CS}^{\rm D8} &= - dt \wedge dx^1 \wedge dx^2 \wedge \frac{R}{3} q_{\rm D8}e^{\frac{3Z}{R}}  \left[\frac{3}{10} \left( dd^\dag \left(f_\star  J_{\rm CY}  \right) + 2 G_0 J_{\rm CY}\right) \wedge \cF^2  + \frac{4}{5}  G_0 g_s \varphi J_{\rm CY}^3 \right] + \dots\\
  & = - dt \wedge dx^1 \wedge dx^2 \wedge \frac{R}{3} e^{\frac{3z_0}{R}}   \left[\frac{3}{5} |G_0| J_{\rm CY} \wedge \cF^2 + \frac{4}{5}  |G_0| g_s \varphi J_{\rm CY}^3 \right] + \dots
  \label{DBICSns}
\end{align}
where we have neglected terms that do not depend on $z_0$, and in the second line we have only kept terms up to order $\cO(g_s^{4/3})$. Out of the two remaining terms, one of them will vanish when integrating over $X_6$, since  $\int_{X_6} \varphi = 0$. The other one finally gives
\be
 Q_{\rm D8}^{\rm BIon, ns} - T_{\rm D8}^{\rm BIon, ns} =  - e^{K/2} \frac{1}{\ell_s^{6}}\int_{\rm X_6} J_{\rm CY} \wedge \cF^2 + \cO(g_s^2) \, .
\label{QTbionnosusy}
\ee
This result is perhaps not very surprising, because it reproduces the result \eqref{QTsmnosusy} of the smearing approximation when curvature corrections are omitted and $\cF$ is a harmonic form. However remember that in the present setup $\cF$ is always non-vanishing, even when the harmonic piece of $\cF$ is set to zero. Therefore, 
\be
2\Delta_{\rm D8}^{\rm Bion} \equiv - e^{K/2} \frac{1}{\ell_s^{6}}\int_{\rm X_6} J_{\rm CY} \wedge \cF^2
\label{QTbionnosusyexp}
\ee
constitutes a correction to the previous result \eqref{QTsmnosusy}. Since a vanishing harmonic piece for $\cF$ is always a choice, there will always be some BIonic membrane whose charge-to-tension ratio will be fixed by the curvature term $2K_a^{(2)} T_{\rm D4}^a$ plus  \eqref{QTbionnosusyexp}. 

One may thus wonder what is the magnitude of $\Delta_{\rm D8}^{\rm Bion}$ compared to $2K_a^{(2)} T_{\rm D4}^a$, as well as its sign. For this notice that \eqref{cfsol} is suppressed as $\cO(g_s^{2/3})$ compared to a harmonic two-form. Therefore $\Delta_{\rm D8}^{\rm Bion}$ gets an relative suppression of $\cO(g_s^{4/3}) \sim {\cal V}_{\rm CY}^{-2/3}$, just like both terms in \eqref{QTsmnosusy}. In other words, $\Delta_{\rm D8}^{\rm Bion}$ and  $2K_a^{(2)} T_{\rm D4}^a$ scale similarly with the string coupling. As for the sign, it will be the result of two competing quantities, since
\be
2\Delta_{\rm D8}^{\rm Bion} =  e^{K/2} \frac{1}{\ell_s^{6}} \int_{X_6}  *_{\rm CY} \cF_2 \wedge \cF_2 -   *_{\rm CY} \cF_1 \wedge \cF_1\, ,
\ee
where $\cF_1 \equiv  \cF^{(1,1)}$ and $\cF_2 \equiv \cF^{(2,0)+(0,2)}$. If we assume \eqref{cfsol} we obtain
\begin{align}
\label{cF1}
\cF_1 & = \frac{i}{2G_0} J_{\rm CY} \cdot \bar{\partial} k = G_0^{-1}J_{\rm CY} \cdot d\left(*_{\rm CY} K - 2 \varphi \im \Omega_{\rm CY} \right) \, , \\
\cF_2 &  = - G_0^{-1} J_{\rm CY} \cdot d\left( 2\varphi \im \Omega_{\rm CY} \right) \, .
\label{cF2}
\end{align}
Intuitively, a $(1,1)$ component of $\cF$ induces D4-brane charge on the BIon worldvolume, and drags it away from the boundary, while a $(2,0)+(0,2)$ component induces anti-D4-brane charge and therefore the opposite effect. So if the integrated norm of $\cF_2$ wins over that of $\cF_1$ the BIonic membrane suffers an additional force that draws it towards the boundary of AdS$_4$, providing a source of instability for the non-supersymmetric vacuum. 

Computing $\Delta_{\rm D8}^{\rm Bion}$ is in general non-trivial, but one may do so for toroidal or toroidal orbifold geometries, where vacua of this sort can be constructed explicitly \cite{Camara:2005dc,Narayan:2010em,Ihl:2006pp}. We have computed its value  in Appendix \ref{ap:torus}, for the particular case of $T^6/ (\Z_2 \times \Z_2)$.  The result is
\be
\Delta_{\rm D8}^{\rm Bion} (T^6/ (\Z_2 \times \Z_2)) = \frac{(8h)^2}{24} \sum_i T_{\rm D4}^i \, ,
\label{finalresult}
\ee
where $T_{\rm D4}^i = e^{K/2} \frac{1}{4} {\cal V}_{T^2_i}$ is the tension of a fractional D4-brane wrapping the $i^{\rm th}$ two-torus. As discussed in Appendix \ref{ap:torus} the factor $(8h)^2$ is related to the O6-plane intersections, while the $1/24$ is reminiscent of \eqref{Kcurv}. It is thus tempting to interpret \eqref{finalresult} as a curvature correction to the BIon action induced by D6-brane intersections. This intuition matches well with the positive sign in \eqref{finalresult} that adds up to the effect of the Calabi--Yau curvature corrections. Together, they mean that the induced charge and tension is negative compared to that of D4-branes, and this drags the BIonic membrane towards the boundary of AdS$_4$. Following our previous discussion, this will induce a non-perturbative instability towards non-supersymmetric vacua with larger values for $|m|$ and less space-time filling D6-branes, until none of the latter remain. A more general analysis of $T^6/ (\Z_2 \times \Z_2)$ and other toroidal orientifolds will be carried in \cite{Casas:2022mnz}. Notice that even if we stabilise the K\"ahler moduli away from the orbifold limit as in \cite{DeWolfe:2005uu,Narayan:2010em}, in the trustable regime the blow-up modes are significantly smaller than the untwisted K\"ahler moduli, and so the sign of $\Delta_{\rm D8}^{\rm BIon}$ remains the same as in \eqref{finalresult}. The excess membrane charge induced by $\Delta_{\rm D8}^{\rm BIon}$ then adds up to the one induced by $K_a^{(2)} T_{\rm D4}^a$ and, as a consequence, a non-perturbative instability is induced for these vacua if D6-branes are present.

%%%%%%%%%%%%%%%%%%%

\section{Conclusions}
\label{s:conclu}

In this paper we have revisited the non-perturbative stability of type IIA $\cN=0$ AdS$_4 \times X_6$ orientifold vacua, where $X_6$ has a Calabi--Yau metric in the smeared-source approximation. For our analysis we have used the results of  \cite{Junghans:2020acz,Marchesano:2020qvg},  which give a description of these backgrounds beyond the Calabi--Yau approximation. Such a description is quite accurate in the large-volume, weak-coupling regime, at least at regions of $X_6$ away from the O6-plane location. However, as already pointed out, we are still working with an approximate solution which will have further corrections at higher orders in the expansion. At such a higher level of accuracy, and specially in non-supersymmetric settings, there will be additional corrections that one should take into account, and which are beyond the scope of the present analysis. 

Given our results, there are several open questions to be addressed. First, we have unveiled a potential decay channel for $\cN=0$ AdS$_4$ vacua with space-time filling D6-branes, triggered by nucleating D8-branes that take the system to a new $\cN=0$ vacuum with larger $|F_0|$ and fewer D6-branes. There are two quantities that determine if this decay channel exist, namely the curvature correction term $K_a^{(2)} T_{\rm D4}^a$ to the D8-brane action and the BIon correction $\Delta_{\rm D8}^{\rm Bion}$ defined in \eqref{QTbionnosusyexp}. The sharpened WGC for membranes \cite{Ooguri:2016pdq} predicts that $K_a^{(2)} T_{\rm D4}^a+\Delta_{\rm D8}^{\rm Bion} > 0$, securing the decay channel. We have shown that this is the case for a simple D6-brane configuration in $X_6 = T^6/ (\Z_2 \times \Z_2)$, and it would be interesting to extend our analysis to other configurations, other toroidal orbifolds and more general Calabi--Yau geometries. In particular it would be interesting to see if the two terms always add up to yield a positive quantity, the key question being how $\Delta_{\rm D8}^{\rm Bion}$ behaves in general. Because $\cF$ is a non-closed but nevertheless quantised two-form, it could be that $\Delta_{\rm D8}^{\rm Bion}$ is determined by the topological data of the problem, as the simple result obtained for toroidal geometries would suggest.

More generally, the instabilities that we have discussed only apply to vacua with space-time filling D6-branes. For instance, the explicit vacua discussed \cite{DeWolfe:2005uu,Narayan:2010em} were based on toroidal orbifolds, but the $H$-flux and $F_0$ quanta were chosen such that no D6-branes were present. For these vacua and others alike, our results find no superextremal membrane that could mediate the decay, since D4-branes saturate a BPS bound in the same sense that they do in the smeared-source approximation analysis. It would be interesting to see if pushing our analysis to the next term in the expansion one could find that $Q_{\rm D4} \neq T_{\rm D4}$ in $\cN=0$ backgrounds, or if some other kind of corrections sourced by supersymmetry-breaking effects creates an imbalance. If not, one may consider more exotic classes of processes where four-form flux is discharged, like decays involve a mixture of bubbles of nothing and D4-brane charge (see e.g. \cite{Bomans:2021ara}) to fully test the sharpened WGC for membranes. 

In any event, we believe that the decay processes that we have studied are interesting per se, and deserve further study. Notice for instance that after bubble nucleation the AdS$_4$ flux dual to the Romans mass is not discharged, as in \cite{Maldacena:1998uz}, but on the contrary it increases. And the same happens with the 4d four-form flux dual to $G_4$. From the 4d viewpoint there is nothing wrong with this fact, as we jump to a new $\cN=0$ vacuum with lower vacuum energy. Indeed, we have argued in \ref{s:insta} that these decays are favourable from the 4d viewpoint, even when we are away from the thin-wall approximation. It would however be interesting to carry a more detailed 4d analysis of this process, as well as to build the explicit 4d solution. Moreover, it would be important to analyse the superextremality of the membranes from a standard 4d viewpoint, like the analysis of the WGC for membranes carried out in \cite{Lanza:2020qmt}.

From the microscopic viewpoint, it would be interesting to see if our computations can be generalised to other string theory settings. Obvious candidates are the class of type IIA orientifold compactifications studied in \cite{Villadoro:2005cu,Banks:2006hg,Cribiori:2021djm}, which share many similar properties with the ones considered in this paper. But one may also consider other compactifications which share key ingredients like scale separation and non-Abelian chiral gauge theories, and see if similar results are obtained. After all, our results hint that $\cN=0$ 4d EFTs with non-trivial gauge sectors are more susceptible to decay to vacua where such gauge sectors are absent. If true in general, this would have deep implications for string theory model building, and probably result into a new branch of implications of the Swampland Programme. 

\bigskip

\bigskip

\centerline{\bf  Acknowledgments}

\vspace*{.5cm}

We would like to thank \'Alvaro Herr\'aez, Jos\'e L. F.~Barb\'on,  I\~naki Garc\'ia-Etxebarria, Thomas Grimm, Luca Martucci, Miguel Montero, Alessandro Tomassiello, \'Angel Uranga, Irene Valenzuela and Cumrun Vafa for useful discussions.  This  work is supported by the Spanish Research Agency (Agencia Estatal de Investigaci\'on) through the grants CEX2020-001007-S and PGC2018-095976-B-C21, funded by MCIN/AEI/10.13039/501100011033 and by ERDF A way of making Europe. The work of D.P. is supported through the FPU grant No. FPU19/04298. J.Q. is supported through the FPU grant No. FPU17/04293. J.Q. would like to thank IPhT Saclay for hospitality during the completion of this work.

%%%%%%%%%%%%%%%%%%%%%%%%%%%%%%%%%%%%%%%%%%%%%%%%%%%%%%%%%%%%%%%%%%%%%%%%%%%%%%%%%%%%%%%%%%%%%%%%%%%%%%%%%%%%%%%%%%%%%%%%%%%%%%%%%%%%%%%%%%%%%%%%%%%%%%%%%%%%%%%%%%%%%%%%%%%%%%%%%%%%%%%%%%%%%%%%%%%%%%%%%%%%%%%%%%%%%%%%%%%%%%%%%%%%%%%%%%%%%%%%%%%%%%%%%%%%%%%%%%%%%%%%

\appendix

%%%%%%%%%%%%%%%%%%%%%%%%%%%%%%%%%%%%%%%%%%%%%%%%%%%%%%%%%%%%%%%%%%%%%%%%%%%%%%%%%%%%%%%%%%%%%%%%%%%%%%%%%%%%%%%% %%%%%%%%%%%%%%%%%%%%%%%%%%%%%%%%%%%%%%%%%%%%%%%%%%%%%%%%

\section{Moduli stabilisation in type IIA Calabi--Yau orientifolds}
\label{ap:CYIIA}

In this appendix we summarise the relevant results \cite{Marchesano:2019hfb} with respect to moduli stabilization in Type IIA Clabi--Yau flux compactifications, adapted to the conventions of \cite{Marchesano:2020qvg} that we use in the main text. In particular, we obtain the expression for the moduli scalar potential and connect the solutions with the relations introduced in section \ref{ss:smeared}.

\subsubsection*{Kähler Potential and Superpotential}

 The type IIA 4d effective theory used in the Calabi--Yau approximation to moduli stabilisation  \cite{DeWolfe:2005uu,Camara:2005dc,Narayan:2010em,Escobar:2018tiu,Escobar:2018rna,Marchesano:2019hfb,Marchesano:2020uqz} was derived in \cite{Grimm:2004ua} by combining the superpotential generated by the RR and NS flux quanta and the classical K\"ahler potential of Calabi--Yau orientifolds without fluxes. In this approach we define the  K\"ahler moduli $T_a=b^a+it^a$ through the complexified K\"ahler form and the two-form basis $l_s^{-2}\omega_a$ of $H^2_-(X_6,\mathbb{Z})$
\begin{equation}
    J_c\equiv B+iJ=T^a\omega_a,\qquad a\in \{1,\cdots,h^{1,1}_-\}\,.
\end{equation}
Similarly, we define the complexified three-form $\Omega_c$ as 
\begin{equation}
    \Omega_c\equiv C_3+i \Im (\mathcal{C}\Omega)\,,
\end{equation}
 where  $\mathcal{C}\equiv e^{-\phi}e^{\frac{1}{2}(K_{cs}-K_K)}$ is a compensator with $K_{cs} = - \log \left(- i\ell_s^{-6} \int_{X_6} \Omega \wedge  \ov \Omega \right)$.

Then, we introduce a symplectic basis $(\alpha_\kappa, \beta^\lambda)$ of $H_3(X_6,\mathbb{Z})$. In this basis, the holomorphic three-form can be written as $\Omega=\mathcal{Z}^\kappa\alpha_\kappa-\mathcal{F}_\lambda \beta^\lambda$. The action of the orientifold projection splits the basis into $\mathcal{R}$-even  and $\mathcal{R}$-odd 3-forms, that is $(\alpha_K,\beta^\Lambda)\in H_+^3(X_6,\mathbb{Z})$ and $(\alpha_\Lambda,\beta^K)\in H_-^3(X_6,\mathbb{Z})$. The complex structure moduli are defined in terms of the intersections of the complexified holomorphic form and the $\mathcal{R}$-odd basis:
\begin{equation}
    N^K=\xi^K+in^K=\ell_s^{-3}\int_{X_6}\Omega_c\wedge\beta^K\,,\qquad U_\Lambda=\xi_\Lambda+iu_\Lambda=\ell_s^{-3}\int_{X_6}\Omega_c\wedge\alpha_\Lambda\,.
\end{equation}
Finally, the following NS integer flux quanta that survive the orientifold projection:
\begin{equation}
    h_K=\frac{1}{\ell_s^2}\int_{X_6}H\wedge \alpha_K\,,\qquad h^\Lambda=-\frac{1}{\ell_s^2}\int_{X_6}H\wedge \beta^\Lambda\,.
\end{equation}

The superpotential and K\"ahler potential are both a sum of the form:
\be
W= W_K + W_Q\, , \qquad K = K_K + K_Q\,.
\label{WK}
\ee
The first term in $W$ is the RR flux-induced superpotential for the K\"ahler moduli \cite{Taylor:1999ii}:
\begin{equation} \label{WT}
\ell_s W_K = - \frac{1}{\ell_s^5} \int_{X_6} {\ov{\bf G}} \wedge e^{J_c} = e_0 + e_a T^a + \frac{1}{2} {\cal K}_{abc} m^a T^b T^c + \frac{m}{6} {\cal K}_{abc} T^a T^b T^c\,,
\end{equation}
where ${\cal K}_{abc} = - \ell_s^{-6} \int_{X_6} \omega_a \wedge \omega_b \wedge \omega_c$ are the Calabi--Yau triple intersection numbers, $\cK = \cK_{abc} t^at^bt^c = 6 {\rm Vol}_{X_6}$ and we have made used of the flux quanta defined in \eqref{RRfluxes}. The second term is a linear superpotential for the complex structure moduli:
\begin{equation} \label{WQ}
\ell_s W_{Q} = -\frac{1}{\ell_s^5} \int_{X_6} \Omega_c \wedge H = h_K N^K  + h^\Lambda U_{\Lambda}\, . 
\end{equation}
Finally, the two pieces that define the K\"ahler potential describe the kinetic terms for K\"ahler moduli
\be
K_K \,  = \, -{\rm log} \left(\frac{i}{6} \CK_{abc} (T^a - \bar{T}^a)(T^b - \bar{T}^b)(T^c - \bar{T}^c) \right) \, = \,  -{\rm log} \left(\frac{4}{3} \cK\right) \, ,
\label{KK}
\ee
and for complex structure moduli
\begin{equation}\label{KQ}
 K_Q = -2 \log \left( \frac{1}{4} \RE({\cal C}{\cal Z}^\Lambda) \IM({\cal C} {\cal F}_\Lambda) - \frac{1}{4} \IM({\cal C} {\cal Z}^K) \RE({\cal C} {\cal F}_K) \right) = -  \log(e^{-4 D})\,.
\end{equation}

\subsubsection*{Vacuum equations and moduli stabilization}

The combination of the RR and NS-fluxes generates a four dimensional F-term scalar potential for the geometric moduli $(t^a,n^k,u_\Lambda)$ and the closed string axions $(b^a,\xi^K,\xi_\Lambda)$. As it is argued in \cite{Bielleman:2015ina,Carta:2016ynn,Herraez:2018vae}, the resulting scalar potential has a bilinear structure that factorises the dependence between axions and saxions
\begin{equation}
    V=\frac{1}{\kappa_4^2}\vec{\rho}^{\, t}\mathbf{Z}\vec{\rho}\,,
\end{equation}
where the matrix $\mathbf{Z}$ encodes all the information regarding the saxions and the vector $\vec{\rho}$ only depends on the flux quanta and the axions. In our case, the explicit form of these objects can be directly derived from the results in \cite{Marchesano:2019hfb}, since our expressions for both the Kähler potential and the superpotential match with those of that paper. We obtain the following axion and flux quanta polynomials:
\begin{subequations}
\begin{align}
    \ell_s\rho_0=&\, e_0+e_ab^a+\frac{1}{2}\mathcal{K}_{abc}m^ab^bb^c+\frac{m}{6}\mathcal{K}_{abc}b^ab^bb^c+h_\mu\xi^\mu\,,\\
    \ell_s\rho_a=&\, e_a+\mathcal{K}_{abc}m^bb^c+\frac{m}{2}\mathcal{K}_{abc}b^bb^c\,,\\
    \ell_s\tilde{\rho}^a=&\, m^a+mb^a\,,\\
    \ell_s\tilde{\rho}=&\, m\,,\\
    \ell_s\hat{\rho}_\mu=&\, h_\mu\,,
\end{align}
\end{subequations}
where we have grouped the complex structure fields and the NS fluxes in a single index for simplicity, i.e, $\xi^\mu=(\xi^K,\xi_\Lambda)$, $u^\mu=(n^K,u_\Lambda)$ and $h_\mu=(h_K,h^\Lambda)$. In this basis, the saxion dependent matrix takes the form
\begin{equation}
\mathbf{Z}=e^K\left(
    \begin{array}{ccccc}
        4 &  &   &   &   \\
        & K^{ab} &   &   &   \\
         &  & \frac{4}{9}\mathcal{K}^2K_{ab}  &   &   \\
          &  &   &  \frac{1}{9}\mathcal{K}^2 &\frac{2}{3}\mathcal{K}u^\mu   \\
           &  &   & \frac{2}{3}\mathcal{K}u^\nu  & K^{\mu\nu}  
\end{array}   \right)\,, 
\end{equation}
with $K_{ab}=\frac{1}{4}\partial_{t^a}\partial_{t^b}K_K$ and $K_{\mu\nu}=\frac{1}{4}\partial_{u^\mu}\partial_{u^\nu}K_Q$.

The solutions to the vacuum equations associated to this potential where found in \cite{Marchesano:2019hfb} to be 
\begin{equation}
    \rho_0=0\,,\qquad\hat{\rho}_\mu=\tilde{\rho}\mathcal{K}(A\partial_{u^\mu}K+\hat{\epsilon}^p_\mu)\,,\qquad \tilde{\rho}^a=B\tilde{\rho}t^a\,,\qquad \rho_a=C\tilde{\rho}\mathcal{K}_a \, .
    \label{eq: rho solutions}
\end{equation}
In the main text we focus on the branch of solutions A1-S1 from \cite{Marchesano:2019hfb}, in which $\hat{\epsilon}_\mu^p=0$ and the parameters $A,B,C$ take the values
\begin{equation}
    A=\frac{1}{15}\,, \qquad B=0\,,\qquad C=\pm\frac{3}{10}\,,
\end{equation}
with the positive choice for the parameter $C$ corresponding to the supersymmetric solution, and the negative choice for the non-supersymmetric one. Substituting this values in the equations for $\rho_0$, $\tilde{\rho}^a$ and $\rho_a$ in  \eqref{eq: rho solutions}, we  recover the relations \eqref{intfluxsm} for $C= \frac{3}{10}$ and \eqref{intfluxsmnosusy} for  $C= -\frac{3}{10}$.

% \begin{equation}\label{KQ}
%  K_Q = -2 \log \left( \frac{1}{4} \RE({\cal C}{\cal Z}_K) \IM({\cal C} {\cal F}^K) - \frac{1}{4} \IM({\cal C} {\cal Z}_\Lambda) \RE({\cal C} {\cal F}^\Lambda) \right) = -  \log(e^{-4 D}),
% \end{equation}

%%%%%%%%%%%%%%%%%%%

\section{10d equations of motion}
\label{ap:10deom}

In this appendix we will discuss how the SUSY \eqref{solutionflux} and the non-SUSY backgrounds \eqref{solutionfluxnnosusy} presented in the main text solve the 10d equations of motion. The Bianchi identities, which also must be satisfied, were discussed in great detail in \cite{Marchesano:2020qvg} for the SUSY case so we will limit ourselves to emphasise the changes the non-SUSY case introduces.

\subsubsection*{RR and NSNS field equations}

In our conventions, the EOM for the field strengths -in the string frame- are
\begin{subequations}
\begin{align}
0&=\mathrm{d}\left(\star_{10} G_{2}\right)+H \wedge \star_{10} G_{4}\, , \\
0 &=\mathrm{d}\left(\star_{10} G_{4}\right)+\cancelto{0}{H \wedge \star_{10} G_{6}}\, , \label{noproblem}\\
0 &=\cancelto{0}{\mathrm{d}\left(\star_{10} G_{6}\right)}\, , \\ 
\label{problem}0 &=\mathrm{d}\left(e^{-2\phi} \star_{10} H\right)+\star_{10} G_{2} \wedge G_{0}+\star_{10} G_{4} \wedge G_{2}+\cancelto{0}{\star_{10} G_{6} \wedge G_{4}}\, ,
\end{align}
\end{subequations} 
where we are taking into account that $G_6=0$. There are also the Einstein and dilaton equations, whose discussion  we leave for the end of the appendix. 

\subsubsection*{Supersymmetric case}

In our approximation, the internal part of the first equation is
\begin{align}
\label{ap1eq1}
0&=\mathrm{d}\left(e^{4A}\star_{\rm CY} G_{2}\right)+e^{4A}H \wedge \star_{\rm CY} G_{4}+\CO\left(g_s\right)=0+\CO\left(g_s\right)\, ,
\end{align}
where we have used that $G_2$ is known up to $\CO\left(g_s\right)$ -see  \eqref{G2sol}. Since the natural scaling of a  $p$-form is $g_s^{-p/3}$, the total error we are making in solving  this equation is  $\CO(g_s^{8/3})$.

The internal part of second equation, at our level of approximation, reads
\begin{align}
0=d\left(e^{4A}\star_{\rm CY} G_4\right)=4g_se^{4A}G_0d\varphi\wedge J_{\text{CY}}+\frac{e^{4A}}{g_s}d\star_{\rm CY} \left(J_{\rm CY}\wedge d\im v\right)+\CO(g_s^2)\, ,
\label{eqg4}
\end{align}
it is more or less straightforward to check that
\begin{align}
\label{auxg4}
\frac{1}{g_s} d\star_{\rm CY} \left(J_{\rm CY}\wedge d\im v\right)=4G_0g_s\star_{\rm CY}\left( J_{\rm CY}\wedge d_c\varphi\right)=-4G_0g_s J_{\rm CY}\wedge d\varphi\, ,
\end{align}
which cancels out the first term of  \eqref{eqg4} and satisfies the equation up to order $ \CO(g_s^{3})$ compared to the natural scaling of a three-form.

It remains to check equation \eqref{problem}, which is the most cumbersome. We will go term by term and write just the internal parts to make the computation clearer. At the level of approximation that we are working the second term in the r.h.s  is 
\begin{align}
e^{4A}\star_6 G_4\wedge G_2=\frac{12}{5}G_0 d\varphi\wedge\im\Omega_{\rm CY}-\frac{3}{5}G_0\star_{\rm CY}G_2+\CO(g_s)\, ,
\end{align}
while the first term reads
\begin{align}\nonumber
d\left(e^{-2\phi+4A}\star _6 H\right)& =d\left[ e^{-2\phi+4A}\star_6\left(\frac{2}{5}G_0g_se^{-A}\re\Omega-\frac{1}{2}d\re\left(\bar{v}\cdot\Omega_{\rm CY}\right)\right)\right]+\CO(g_s)\\
& = \frac{2 }{5}G_0g_sd\left(e^{-2\phi+3A}\im\Omega\right)-\frac{e^{-2\phi+4A}}{2}d\star_6d\re\left(\bar{v}\cdot\Omega_{\rm CY}\right)+\CO(g_s)\, .
\label{ap:divided}
\end{align}
The first contribution to \eqref{ap:divided} can then be rewritten as
\begin{align}
\frac{2 }{5}G_0g_sd\left(e^{-2\phi+3A}\im\Omega\right)=\frac{2G_0}{5}\left(4d\varphi\im\Omega_{\rm CY}-\star_{\rm CY}G_2\right)+\CO(g_s)\, ,
\end{align}
whereas for the second contribution, a long calculation shows that
\begin{align}
-\frac{e^{-2\phi+4A}}{2}d\star_6d\re\left(\bar{v}\cdot\Omega_{\rm CY}\right)=-4G_0 d\varphi\wedge\im\Omega_{\rm CY}+\CO(g_s)\, .
\end{align}
Finally, putting everything together \eqref{problem} reduces to 
\begin{align}
0=\left(\frac{12}{5}+\frac{8}{5}-4\right)e^{4A}G_0d\varphi\wedge\im\Omega_{\rm CY}+\left(-\frac{2}{5}-\frac{3}{5}+1\right)e^{4A}G_0\star_{\rm CY}G_2+\CO(g_s)\, ,
\end{align}
which,  as a $4$-form equation, we are solving it with an error  $\CO(g_s^{7/3})$.

\subsubsection*{Non-supersymmetric case}

In the non-SUSY solution, only the fields $H$ and $G_4$ change, so it is enough to check the equations involving these quantities.

Let us start by the Bianchi identities, which we ignored in the previous section. To start with we can look at
\begin{align}
\label{bianH}
d G_4= G_2\wedge H\, .
\end{align}
The changes in $G_4^{\text{non-SUSY}}$ appear in the harmonic and the closed parts, so the LHS is the same as the  $G_4^{\text{SUSY}}$. The changes in $H^{\text{non-SUSY}}$ are of order $\CO(g_s^{7/3})$, giving a  contribution beyond the order at which \eqref{bianH} is being solved: we can ignore them and recover the RHS of the SUSY solution as well. The other BIs which could be sensitive to the non-SUSY novelties are  $dG_2$ and $dH$. For both of them, the changes appear beyond the order of approximation in which they are being solved, so we can just neglect them.

Regarding the equations of motion, for $G_4$  the internal part now reads
\begin{align}
d\left(e^{4A}\star_{\rm CY} G_4\right)=-\frac{24}{5} g_se^{4A}G_0d\varphi\wedge J_{\text{CY}}-\frac{6e^{4A}}{5g_s}d\star_{\rm CY} \left(J_{\rm CY}\wedge d\im v\right)+\CO(g_s^2)=0+\CO(g_s^2)\, ,
\end{align}
where we have used \eqref{auxg4}. As in the SUSY case, it is solved at total order $\CO(g_s^3)$. 

Finally, the equation for $H$ is again the most tedious. Following the reasoning of the previous section, we will directly write each of the contributions to the internal part. On the one side
\begin{align}
e^{4A}\star_6 G_4\wedge G_2=-\frac{12}{5}G_0 d\varphi\wedge\im\Omega_{\rm CY}+\frac{3}{5}G_0\star_{\rm CY}G_2+\CO(g_s)\, ,
\end{align}
on the other side
\begin{align}
d\left(e^{-2\phi+4A}\star _6 H\right)=\frac{12}{5}G_0d\varphi\wedge\im\Omega_{\rm CY}-\frac{8}{5}G_0\star_{\rm CY}G_2+\mathcal{O}(g_s)\, ,
\end{align}
and \eqref{problem} reduces to
\begin{align}
0=\left(-\frac{12}{5}+\frac{12}{5}\right)e^{4A}G_0d\varphi\wedge\im\Omega_{\rm CY}+\left(-\frac{8}{5}+\frac{3}{5}+1\right)e^{4A}G_0\star_{\rm CY}G_2+\CO(g_s)\, ,
\end{align}
which is again solved  at total order $\CO(g_s^{7/3})$.

\subsubsection*{Einstein and dilaton equations}

To show how our expressions satisfy these two last constraints, we will use the results derived in \cite{Junghans:2020acz}, focusing again on the changes introduced by the non-SUSY case. At leading order the equations evaluated for the non-SUSY solution coincide with the equations evaluated in the SUSY background, so they are satisfied in the first case provided they are solved in the second case -as it happens-. When the changes come into play, they do it at least at order $|F_4|^2\sim e^{-2\phi}|H|^2\sim \CO(g_s^{4/3})$. Nevertheless, to solve the equations at this order, we need to consider terms in $e^{A}$ and $e^{-\phi}$ which are beyond our approximation. %\footnote{Terms like $e^{-2\phi}\partial e^{A}$ which at order $\CO\left( g_s^{4/3}\right)$ contribute with  $g_s^{-2} (\partial e^A|_{\CO\left( g_s^{10/3}\right)})$.}
  In other words, the modifications introduced in the non-SUSY case are seen by the Einstein and dilaton equations at the next order in the expansion.

%%%%%%%%%%%%%%%%%%%%%%

\section{DBI computation}
\label{ap:dbi}

The BIonic D8-brane system of section \ref{s:bion} is defined by the profile \eqref{BIonrel} for the transverse D8-brane position. In this appendix we check that this relation fulfils the basic requirement of a BPS condition, in the sense that it linearises the DBI action of the D8-brane, at least at the level of approximation at which we work in the main text. 

The DBI action of a D8-brane wrapping $X_6$ is given by 
\be
 S_{\rm DBI}^{\rm D8} = - \frac{2\pi}{\ell_s^9} \int dt dx^1 dx^2 \int_{X_6} d^6 \xi   e^{3A - \phi} e^{\frac{3Z}{R}} \sqrt{\det \left(g_{ab} + \p_aZ\p_bZ +\cF_{ab} \right) } \, ,
 \label{ap:DBID8}
\ee
where the D8-brane transverse position $Z$ is seen as a function on $X_6$. For BPS configurations the integrand simplifies, in the sense that the square root linearises and corresponds to integrating a six-form over $X_6$. To see how this happens for the BIon configuration, let us use the matrix determinant lemma to rewrite things as
\be
\det \left(g_{ab} + \p_aZ\p_bZ +\cF_{ab} \right)  =  \det g \, \det \left(\II + g^{-1} \cF\right) \left( 1 + \p Z \cdot (g+\cF)^{-1} \cdot \p Z\right)\, .
\label{inidet}
\ee
Then using that $\cF$ is antisymmetric one can deduce that
\begin{align}
 \det \left(\II + g^{-1} \cF\right) = 1 - \frac{t_2}{2} + \frac{t_2^2}{8} -\frac{t_4}{4} + \frac{\det  \cF}{\det g}\, ,
\end{align}
where  $t_k = \Tr\, g^{-1} \cF^k$. Using in addition the Woodbury matrix identity we obtain 
\be
\p Z \cdot (g+\cF)^{-1} \cdot \p Z = \p Z \cdot \sum_{k=0}^\infty \left(g^{-1} \cF\right)^{2k} g^{-1} \cdot \p Z\, .
\ee

One may then combine all these expressions to compute \eqref{inidet}. Recall however that our unsmeared background description is only accurate below $\cO(g_s^2)$ corrections in the $g_s$ expansion. As pointed out in  \cite{Junghans:2020acz,Marchesano:2020qvg} a flux of the form \eqref{cfsol} is suppressed as $\cO(g_s^{3/2})$ compared to a harmonic two-form and, because of \eqref{BIonrel}, the same suppression holds for $\p Z$. This means that we are only interested in terms up to quadratic order in the worldvolume flux or $\p Z$ in the DBI action, or equivalently up to quartic order in \eqref{inidet}. That is, we are interested in computing the following terms
\be
\left(1 - \oh \Tr \tilde\cF^2 \right) \left(1 + (\p Z)^2\right)   + \frac{1}{8} \left( \Tr \tilde\cF^2\right)^2 -\frac{1}{4}  \Tr \tilde\cF^4 - \left( \p Z\cdot  \tilde \cF \right)^2 \, ,
\ee
where $\tilde \cF \equiv g_{\rm CY}^{-1} \cF$, and $(\p Z)^2 = g_{\rm CY}^{ab} \p_a Z \p_b Z$, etc. To proceed we split the worldvolume flux as in section \ref{ss:bionnosusy}
\be
\tilde \cF_1 \equiv g^{-1}  \cF^{(1,1)} \, , \qquad \tilde \cF_2 \equiv g^{-1} \cF^{(2,0)+(0,2)} \, ,
\ee
assuming that $\cF^{(1,1)}$ is primitive, and use the following identity
\be
\Tr \tilde\cF^4 = \frac{1}{4} \left(\Tr \tilde \cF^2 \right)^2 +  \left(\Tr \tilde \cF_1^2 \right)\left(\Tr \tilde \cF_2^2 \right) + 4 \Tr \left([\tilde \cF_1, \tilde \cF_2]^2\right) \, ,
\ee
to arrive to
\be
\left( 1 - \frac{1}{4} \Tr \tilde\cF^2 + \oh (\p Z)^2 \right)^2 -  \frac{1}{4}\left((\p Z)^2  \Tr \tilde\cF^2 +  \Tr \tilde \cF_1^2 \, \Tr \tilde \cF_2^2 +  (\p Z)^4 \right)  -\Tr \left([\tilde \cF_1, \tilde \cF_2]^2\right) - \left( \p Z\cdot  \tilde \cF \right)^2\, .
\ee
Finally, one can see that \eqref{BIonrel} and primitivity imply that
\be
 (\p Z)^2 = -  \Tr \tilde\cF_2^2\, , \qquad \left( \p Z\cdot  \tilde \cF \right)^2 = \left( \p Z\cdot  \tilde \cF_1 \right)^2 = - \Tr \left([\tilde \cF_1, \tilde \cF_2]^2\right)\, ,
\ee
and so we are left with
\be
\left( 1 - \frac{1}{4} \Tr \tilde\cF^2 + \oh (\p Z)^2 \right)^2 = \left( 1 - \frac{1}{4} \Tr \left(\tilde\cF_1^2  - \tilde\cF_2^2\right) + (\p Z)^2 \right)^2\, .
\ee
When plugged into \eqref{ap:DBID8} this translates into
\be
 S_{\rm DBI}^{\rm D8} = - \frac{2\pi}{\ell_s^9} \int dt dx^1 dx^2  g_s^{-1} e^{\frac{3z_0}{R}} \int_{X_6} \left[-\frac{1}{6}J_{\rm CY}^3 + \oh J_{\rm CY} \wedge \cF^2 + *_{\rm CY} dZ \wedge dZ + \cO(g_s^{4/3}) \right]
 \label{ap:DBID8fin}
\ee
where we used that in our approximation $\cF_1 \equiv \cF^{(1,1)}$ is a primitive (1,1)-from, and as a result $-\oh \Tr \tilde \cF_1^2 d{\rm vol}_{X_6} = \star_{\rm CY} \cF_1 \wedge \cF_1 = J_{\rm CY} \wedge \cF_1 \wedge \cF_1$. Finally, we have expanded $e^{3A-\phi} = g_s^{-1} + \cO(g_s)$ and $e^{\frac{3Z}{R}} = e^{\frac{3z_0}{R}}\left(1 - \frac{12 \ell_s}{|m|R}  \varphi \right) + \cO(g_s^{8/3})$, and used that $\int_{X_6} \varphi = 0$.

%%%%%%%%%%%%%%%%%%%%%%w

\section{BIonic strings and SU(4) instantons}
\label{ap:IIBion}

The BIonic solution found in section \ref{s:bion} is not unique of type IIA flux compactifications. It can also be found when one wraps a D7-brane on the whole internal manifold of type IIB warped Calabi--Yau compactifications with background three-form fluxes. The advantage of this type IIB setup compared to the type IIA one considered in the main text is two-fold: {\it i)} we know the exact 10d background and  {\it ii)} we can directly connect it to the Abelian $SU(4)$ instanton solutions that define Donaldson--Thomas theory \cite{Donaldson:1996kp}.

\subsubsection*{IIB BIonic strings}

Let us consider a type IIB warped Calabi--Yau compactification, namely a metric background of the form
\be
ds^2 = e^{2A}ds^2_{\pr^{1,3}} + e^{-2A} ds^2_{X_6}\, ,
\ee
where $X_6$ is endowed with a Calabi--Yau metric. On top of it we can add background fluxes $H$ and $F_3$ which are quantised harmonic three-forms of $X_6$ sourcing the warp factor. Let us consider the case in which $\ell_s^{-2} [H]$ is Poincar\'e dual to a three-cycle class with a special Lagrangian representative $\Pi$ calibrated by $\im \Omega_{\rm CY}$. That is:
\be
 \ell_s^{-2} [H] = {\rm P.D.} [\Pi]  = \ell_s^{-3} \d (\Pi)\, ,
\ee
where $\d (\Pi)$ is the bump delta-function of $X_6$ with support in $\Pi$. 

We now wrap a D7-brane on the internal six-dimensional space, as in \cite[section 6]{Evslin:2007ti}, and extended along $(t,x^1,0,0)$. The Freed--Witten anomaly induced by the $H$-flux can be cured by a D5-brane wrapping $-\Pi$, extended along $(t, x^1,0, x^3>0)$ and ending on the D7-brane. This configuration describes a 4d string to which a 4d membrane is attached. Microscopically this is due to the Freed--Witten anomaly. Macroscopically it as a result of the type IIB axion $C_0$ gaining an F-term axion-monodromy potential generated by the internal $H$-flux \cite{BerasaluceGonzalez:2012zn,Marchesano:2014mla,Blumenhagen:2014gta}. 

The Bianchi identity for the D7-brane worldvolume flux reads
\be
d\cF =  H - \ell_s^{-1}  \d (\Pi)\, ,
\label{ap:BIF}
\ee
and finding its solution works as in \cite[section 5]{Marchesano:2020qvg}, see also \cite[section 3.4]{Hitchin:1999fh}. We have that
\be
\ell_s^{-1}\cF = d^{\dag}_{\rm CY} K = - J_{\rm CY} \cdot d \left( \hat\varphi \im \Omega_{\rm CY} - *_{\rm CY} K\right) \, ,
\ee
up to a harmonic piece. Here the function $\hat{\varphi}$ satisfies $\int_{X_6} \hat{\varphi} =0$ and 
\be
\Delta_{\rm CY}  \hat \varphi = \left(\frac{{\cal V}_{\Pi}}{{\cal V}_{\rm CY}} - \delta^{(3)}_{\Pi}\right) \, , \qquad \delta_{\Pi}^{(3)} = *_{\rm CY} \left[ \im \Omega \wedge \d (\Pi)\right]\, ,
\ee
while the three-form current $K$ is defined as in \eqref{formK} with the replacement $\varphi \to \hat{\varphi}/4$. The main difference with respect to the type IIA solution is that this one is exact. The 10d BPS configurations is therefore described by a BIon solution with profile
\be
*_{\rm CY} dX^3 =  \im \Omega_{\rm CY} \wedge \cF\, ,
\ee
from where we deduce that $X^3 = -  \ell_s \hat{\varphi}$. This would correspond to a DBI action such that
\bea\nonumber
S_\text{DBI}^{\rm D7} &= & - \frac{2\pi}{\ell_s^9} \int dt dx^1 dx^2  g_s^{-1}  \int_{X_6} e^{2A} \sqrt{\det \left(g_{ab} + e^{2A} \p_a X^3 \p_b X^3  + \mathcal{F}_{ab}\right)} \\ 
& =&- \frac{2\pi}{\ell_s^9} \int dt dx^1 dx^2  g_s^{-1} \int_{X_6}  - \frac{e^{-4A}}{6}J^3_{\rm CY} + \frac{1}{2} \cF \wedge \cF \wedge J_{\rm CY}  +  *_{\rm CY} dX^3 \wedge  dX^3\, ,
\label{DBI}
\eea
as would follow from the results of \cite{Evslin:2007ti}. 

Besides being an exact solution, the D7-brane setup has the interesting feature that the transverse space to the D7 is given by $\pr\times S^1$. As a result one is able to relate the D7 BIon system to a gauge configuration that is defined on $\pr\times S^1 \times X_6$. The natural object where such a gauge theory is defined is a  D9-brane dual to the BIonic D7-brane. As we will now discuss, this construction leads us directly to the setup where Donaldson--Thomas theory is defined. 

\subsubsection*{The Donaldson--Thomas setup}

In a Calabi--Yau four-fold $X_8$ we can define a complex star operator $\star$ that maps a $(0,q)$-form $\alpha$ to a $(0,4-q)$-form $\star \a$ such that 
\be
\alpha \wedge \star \alpha =\frac{1}{4} |\alpha|^2 \bar{\Omega}
\ee
where $\Omega$ is the holomorphic four-form of $X_8$, normalised such that $\Omega \wedge \bar{\Omega} = 16 \, d{\rm vol}_{X_8}$. It turns out that $\star$ maps $(0,2)$-forms to $(0,2)$-forms, and that $\star^2 =1$. One can then define two eigenspaces of $(0,2)$-forms such that $\star \alpha_\pm = \pm \a_\pm$. In particular, one may take the $(0,2)$-component of a real non-Abelian gauge flux $F$ on $X_8$ and demand that $\star F^{0,2} = - F^{0,2}$, or in other words that $F^{0,2}_+ =0$. This is one of the conditions of Donaldson--Thomas $SU(4)$ instanton equations  \cite{Donaldson:1996kp}, that read
\begin{subequations}
\label{DT}
\begin{align}
\label{DT1}
       F_+^{0,2}  & = 0\, ,\\
       F \wedge J^3 & = 0 \, .
        \label{DT2}
\end{align}
\end{subequations}

To connect with the more familiar Hodge star operator $*$, one can use that, when acting on $(0,q)$-forms, $\bar{*} = \frac{1}{4}  \Omega \wedge \star$ \cite{Baulieu:1997jx}. Therefore we deduce that
\be
* F^{0,2}_\pm = \pm \frac{1}{4} \bar \Omega \wedge F^{2,0}_\pm\, .
\label{DTH}
\ee
From here we deduce that $ F_\pm^{0,2} =0$ is equivalent to
\begin{subequations}
\label{DTre}
\begin{align}
\label{DTrere}
*  \re F^{0,2} & = \pm\frac{1}{4} \re \Omega \wedge F\, , \\
*  \im F^{0,2} & = \mp \frac{1}{4}\im \Omega \wedge F \ \implies \ F \wedge F \wedge \im \Omega = 0\, .
\label{DTreim}
\end{align}
\end{subequations}
and also implies
\be
\Tr \left(\re F^{0,2} \wedge * \re F^{0,2}\right) = \frac{1}{4} \Tr \left(\re F^{0,2}_+ \wedge \re F^{0,2}_+  - \re F^{0,2}_- \wedge \re F^{0,2}_- \right) \wedge \re \Omega\, .
\label{splitFpm}
\ee

\subsubsection*{The dictionary}

To connect with the D7 BIon configuration, we consider the Donaldson--Thomas equations for an Abelian gauge theory in the following Calabi--Yau background
\be
  \pr  \times S^1 \times X_6\, ,
\label{DTsetup}
\ee
with complex coordinates $\{ \omega =  x + i\theta, z^1, z^2, z^3\}$ and holomorphic four-form
\be
\Omega_4 = \left(dx + i d\theta\right) \wedge \Omega_3 \, .
\ee
We now consider a gauge field strength of the form 
\be
\cF = \cF_{X_6} + \cF_{\rm Bion}\, ,
\label{fluxDT}
\ee
where $\cF_{X_6}$ is a two-form on $X_6$ and 
\be
\cF_{\rm Bion} = F_{xi} \, dx \wedge dz^i + {\rm c.c.}
\ee
so that there is no component of the flux along $d\theta$, and as a result $\cF^4 = 0$. 

The dictionary with the D7 BIon configuration can then be done by simple dimensional reduction along $\pr \times S^1$. After that, we recover a gauge theory on $X_6$ with gauge field strength $\cF_{X_6}$ and a non-trivial profile for the transverse position field $X$, seen as a function on $X_6$
\be
\p X = - F_{xi} dz^i\, .
\ee
Notice that
\be
\cF_{\rm BIon} =  dZ \wedge dx =  \frac{1}{2}\left( \p X + \bar{\p} X \right) \wedge \left( d\omega + d\bar{\omega}\right)  \implies \cF_{\rm BIon}^{0,2} = - \frac{1}{2} d\bar{\omega} \wedge  \bar{\p} Z\, .
\ee
Therefore to satisfy \eqref{DT1} we need to impose 
\be
d\bar{\omega} \wedge  \bar{\p} X = - \oh *_4 \left(\bar{\Omega}_4 \wedge \cF_{X_6}\right) \implies   \bar{\p} X  = \frac{i}{2}  *_{X_6}  \left( \bar{\Omega}_3 \wedge \cF_{X_6}\right)\, ,
\ee
from where we deduce the following relations
\bea
\label{BIon1}
*_{X_6} dZ & = &   \im \Omega_3 \wedge \cF_{X_6} \, ,\\
*_{X_6} d^c Z  & = &   \re \Omega_3 \wedge \cF_{X_6}\, .
\label{BIon2}
\eea
Eq.\eqref{BIon1} corresponds to the BIon equation of section \ref{s:bion}, while \eqref{BIon2} looks like a new, independent equation. In principle we would expect that it also satisfied by the BIon solution, and so it would be interesting to understand its implications. Notice that we can also translate \eqref{DT1} into the condition 
\be
\cF_{X_6}^{0,2} = \frac{1}{8} *_4 \left( \bar{\Omega}_4 \wedge \p X \wedge d\omega\right) \implies \cF_{X_6}^{0,2} = -\frac{i}{4} *_3  \left(\bar{\Omega}_3 \wedge \p X \right) \, ,
\ee
which in turn implies
\bea
\label{BIon3}
\re \cF_{X_6}^{2,0}  & = & - \frac{1}{4} *_3  \left(dX \wedge\im \Omega_3 \right) = \frac{1}{4} *_3  d\left(\hat{\varphi} \im \Omega_3 \right) \, ,\\
\im \cF_{X_6}^{2,0}  & = & - \frac{1}{4} *_3  \left(dX \wedge\re \Omega_3 \right) = \frac{1}{4} *_3  d \left(\hat{\varphi} \re \Omega_3 \right) \, .
\label{BIon4}
\eea
Eq.\eqref{BIon3} corresponds to \eqref{cF2} adapted to this setup, while \eqref{BIon4} is equivalent to \eqref{BIon2}. Finally, imposing \eqref{DT2} amounts to require that $\cF_{X_6}$ is primitive, as the BIon solution fulfils. 

The relation between the solutions to the Bianchi identity of the form \eqref{ap:BIF} and the Abelian SU(4) instanton equations of \cite{Donaldson:1996kp} was already pointed out in \cite[section 3.4]{Hitchin:1999fh}. We find it quite amusing that a BIonic D7-brane and the corresponding worldvolume flux on a D9-brane give a neat physical realisation of this correspondence. It would be interesting to understand if this description has any implications for the theory of invariants developed in \cite{Donaldson:1996kp}.

%%%%%%%%%%%%%%%%%%%

\section{A toroidal orbifold example}
\label{ap:torus}

In this appendix we compute the BIon correction to the D8/D6-system tension  $\Delta_{\rm D8}^{\rm Bion}$ defined in \eqref{QTbionnosusyexp}, for the particular geometry $X_6=T^6/(\mathbb{Z}_2\times \mathbb{Z}_2)$ in the orbifold limit and a specific D6-brane configuration. This case was already analysed in \cite{Marchesano:2020qvg} whose notation we follow up to small modifications. The Calabi--Yau structure is defined as
\bea
J_{\rm CY} & = & \ell_s^2 t_i dx^i \wedge dy^i \, \label{eq: torus J},\\
\re \Om_{\rm CY} & = &  \rho \left(\tau_1\tau_2\tau_3 \b^0 - \tau_1 \b^1 - \tau_2 \b^2 - \tau_3 \b^3 \right) \, ,\\
\im \Om_{\rm CY} & = &   \rho \left( \a_0 - \tau_2\tau_3 \a_1 - \tau_1\tau_3 \a_2 - \tau_1\tau_2 \a_3 \right) \, ,
\eea
where $x^i, y^i$ are period-one coordinates and 
\be
t^i  =  4\pi^2\hat{R}_{x^i} \hat{R}_{y^i}\, , \qquad \tau_i = \frac{\hat{R}_{y^i}}{\hat{R}_{x^i}}\, , \qquad \rho =  8\pi^3 \sqrt{\frac{t_1 t_2 t_3}{\tau_1\tau_2\tau_3}} =  8\pi^3 \hat{R}_{x^1}\hat{R}_{x^2}\hat{R}_{x^3 }\, ,
\ee
with $\ell_s \hat{R}_{x^i}$ and $\ell_s\hat{R}_{y^i}$ the radii of the compact dimensions and we have the following basis of bulk three-forms:
\bea\nonumber
\a_0 = dx^1 \wedge dx^2 \wedge dx^3\, , & \quad & \b^0 = dy^1 \wedge dy^2 \wedge dy^3 \, ,\\ \nonumber
\a_1 = dx^1 \wedge dy^2 \wedge dy^3\, , & \quad & \b^1 = dy^1 \wedge dx^2 \wedge dx^3 \, ,\\ \nonumber
\a_2 = dy^1 \wedge dx^2 \wedge dy^3\, , & \quad & \b^2 = dx^1 \wedge dy^2 \wedge dx^3 \, ,\\ \nonumber
\a_3 = dy^1 \wedge dy^2 \wedge dx^3\, , & \quad & \b^3 = dx^1 \wedge dx^2 \wedge dy^3 \, .
\eea

The BIon worldvolume flux can be derived from \eqref{cfsol} and the results in  \cite[section 6.2]{Marchesano:2020qvg}, generalised to the case where the torus radii are not equal. We obtain
\begin{align}
\cF =& \, \frac{\ell_s}{m} d^\dag_{\text{CY}} K=h \ell_s^4 \star_{\text{CY}} d \star_{\text{CY}}\left(B_0\beta^0-B_1\beta^1-B_2\beta^2-B_3\beta^3\right)\nonumber\\ =&\, h \ell_s^2 \sum_{\eta}\sum_{\vec{0}\neq\vec{n}\in \IZ^3}  2\pi i\left[\frac{e^{i 2\pi  \vec{n}\cdot \left[(y_1, y_2,y_3)+\vec{\eta}\right]}}{ |\vec{n}_{y1,y2,y3}|^2} \left(\frac{n_1}{\tau_1 t_1} dy_2\wedge dy_3-\frac{n_2}{\tau_2 t_2}dy_1\wedge dy_3+\frac{n_3}{\tau_3 t_3}dy_1\wedge dy_2\right)\right. \nonumber\\ &\left. \textcolor{black}{-} \frac{ e^{i 2\pi  \vec{n}\cdot \left[(y_1,x_2,x_3)+\vec{\eta}\right]}}{|\vec{n}_{y1,x2,x3}|^2}\left(\frac{n_1}{\tau_1 t_1} dx_2\wedge dx_3-\frac{n_2\tau_2}{t_2} dy_1\wedge dx_3+\frac{n_3\tau_3}{t_3} dy_1\wedge dx_2\right)\right. \nonumber\\ &\left. \textcolor{black}{-}\frac{e^{i 2\pi  \vec{n}\cdot\left[ (x_1, y_2, x_3)+\vec{\eta}\right]}}{|\vec{n}_{x1,y2,x3}|^2}  \left(\frac{n_1 \tau_1}{t_1} dy_2\wedge dx_3 -\frac{n_2}{t_2\tau_2} dx_1\wedge dx_3+\frac{n_3\tau_3}{t_3} dx_1\wedge dy_2\right)\right. \nonumber\\ &\left. \textcolor{black}{-}\frac{ e^{i 2\pi  \vec{n}\cdot \left[(x_1,x_2,y_3)+\vec{\eta}\right]}}{|\vec{n}_{x1,x2,y3}|^2} \left(\frac{n_1\tau_1}{t_1} dx_2\wedge dy_3 -\frac{n_2\tau_2}{t_2} dx_1\wedge dy_3+\frac{n_3}{\tau_3 t_3} dx_1\wedge dx_2\right)\right]\, ,
\end{align}
where the functions $B_i$ are defined as in \cite[eq.(6.18)]{Marchesano:2020qvg}, $\vec{\eta}$ has entries that are either $0$ or $1/2$ and $|\vec{n}_{x1,x2,y3}|^2=\left(n_1/\hat{R}_{x^1} \right)^2+\left(n_2/\hat{R}_{x^2}\right)^2+\left(n_3/\hat{R}_{y^3}\right)^2$ and similar for the other indices $\{x_i, y_j\}$. 

Using the above expression together with \eqref{eq: torus J}, we arrive to
\begin{align}
\mathcal{F}^2\wedge J_\cy&=8\pi^2h^2 \ell_s^6\sum_{\vec\eta,\vec\eta'} \sum_{\vec{0}\neq\vec{n}\in \IZ^3}\sum_{\vec{0}\neq\vec{m}\in \IZ^3} e^{i2\pi\left(\vec{n}\cdot \vec{\eta}+\vec{m}\cdot \vec{\eta'}\right)}\left[\left( \frac{\tau_1^2}{t_1} \frac{e^{i2\pi\vec{n}\cdot\left(x_1,x_2,y_3\right)+i2\pi\vec{m}\cdot\left(x_1,y_2,x_3\right)}}{|\vec{n}|^2_{x1,x2,y3}|\vec{m}|^2_{x1,y2,x3}}\right. \right.\nonumber \\ &  \left. \left.+\frac{1}{t_1\tau_1^2}\frac{e^{i2\pi\vec{m}\cdot\left(y_1,x_2,x_3\right)+i2\pi\vec{n}\cdot\left(y_1,y_2,y_3\right)}}{|\vec{n}|^2_{y1,x2,x3}|\vec{m}|^2_{y1,y2,y3}}\right)m_1n_1+\dots \right]%\nonumber\\
 dx_1\wedge dx_2\wedge dx_3\wedge dy_1\wedge dy_2\wedge dy_3\, .
\label{gorda}
\end{align}

Now we would like to compute $\int\mathcal{F}^2\wedge J_\cy$ integrating each piece of \eqref{gorda}, but to perform these integrals we first need to regularise them. We do so by smearing the O6-plane over a region of radius $\sim \ell_s$, which is the region of $X_6$ where the supergravity approximation cannot be trusted. 
In practice this corresponds to a truncation of the summation over the Fourier modes labelled by $\vec{n}$ and $\vec{m}$. This allows us to interchange the order between summation and integration. We then take the limit when the cut-off of the sum $N$ diverges, returning to our original system with a localised O6-plane. For the first piece of the integral we obtain 
\begin{align}
   & \lim_{N\to\infty} \sum_{\substack{\{\vec{0}\neq\vec{n}\in \IZ^3\mid |\vec{n}|\leq N\}\\ \{\vec{0}\neq\vec{m}\in \IZ^3||\vec{n}|\leq N\}}}\sum_{\vec\eta,\vec\eta'} e^{i2\pi\left(\vec{n}\cdot \vec{\eta}+\vec{m}\cdot \vec{\eta'}\right)}\int_{T^6}\frac{\tau_1^2}{t_1} \frac{e^{i2\pi\vec{n}\cdot\left(x_1,x_2,y_3\right)+i2\pi\vec{m}\cdot\left(x_1,y_2,x_3\right)}}{|\vec{n}|^2_{x1,x2,y3}|\vec{m}|^2_{x1,y2,x3}}m_1n_1\nonumber\\
    =& \lim_{N\to\infty} \sum_{\substack{\{\vec{0}\neq\vec{n}\in \IZ^3\mid |\vec{n}|\leq N\}\\ \{\vec{0}\neq\vec{m}\in \IZ^3||\vec{n}|\leq N\}}}\sum_{\vec\eta,\vec\eta'} e^{i2\pi\left(\vec{n}\cdot \vec{\eta}+\vec{m}\cdot \vec{\eta'}\right)}\frac{m_1n_1\tau_1^2}{t_1|\vec{n}|^2_{x1,x2,y3}|\vec{m}|^2_{x1,y2,x3}} \delta(n_1+m_1)\delta(n_2)\delta(n_3)\delta(m_2)\delta(m_3)\nonumber\\
    =& -\sum_{\vec{0}\neq n_1\in \IZ}\sum_{\vec\eta,\vec\eta'} e^{i2\pi n_1\left( \eta_1-\eta_1'\right)}\frac{n_1^2 \hat{R}_{x^1}^4\tau_1^2}{n_1^4 t_1}\,.
\end{align}

Repeating a similar process for all the contributions in \eqref{gorda} and adding them together we conclude
\begin{align}
    \frac{1}{\ell_s^6}\int_{X_6}  \mathcal{F}^2\wedge J_{CY}=&-\frac{1}{2\pi^2} h^2(t_1+t_2+t_2)\sum_{n=1}^{\infty}\frac{1}{n^2}\sum_{\vec{\eta},\vec{\eta}'}e^{i2\pi n(\eta_1-\eta_1')}\nonumber\\
    =&-\frac{16}{\pi^2} h^2(t_1+t_2+t_3)\sum_{n=1}^\infty\left(\frac{1}{n^2}+\frac{(-1)^n}{n^2}\right)\nonumber\\
    =&-\frac{16h^2}{12}\,  (t_1+t_2+t_3)   \, ,
\end{align}
where we have taken into account that the integration space of $X_6 = T^6/(\mathbb{Z}_2 \times \mathbb{Z}_2)$ is $1/4$ of that of $T^6$. Notice that the result goes like the square of the number of D6-branes, and therefore as their pairwise intersection. Indeed, in this case the $\Pi_{\rm O6}$ is composed of $4 \times 8$ different three-cycles, with each group of 8 three-cycles wrapping the same class of $T^6$:
\be
\ell_s^3 \text{P.D.} [\Pi_{\rm O6}] = 8  [\beta^0] - 8[\beta^1] - 8[\beta^2] - 8[\beta^3]\, .
\ee
Two different three-cycles intersect over one one-cycle, so there are 64 intersections arising from each pair of classes. Finally, because there are  $h$ D6-branes wrapping $\Pi_{\rm O6}$, this factor increases to $(8h)^2$. It would be interesting to generalise this result to more involved D6-brane configurations, and in particular those where they do not lie on top of O6-planes. 

%%%%%%%%%%%%%%%%%%%%%%

%\newpage

%\addcontentsline{toc}{section}{References}
\bibliographystyle{JHEP2015}
\bibliography{papers}

\providecommand{\href}[2]{#2}\begingroup\raggedright\begin{thebibliography}{10}

\bibitem{Vafa:2005ui}
C.~Vafa, \emph{{The String landscape and the swampland}},
  \href{https://arxiv.org/abs/hep-th/0509212}{{\ttfamily hep-th/0509212}}.

\bibitem{Brennan:2017rbf}
T.~D. Brennan, F.~Carta and C.~Vafa, \emph{{The String Landscape, the
  Swampland, and the Missing Corner}},
  \href{https://doi.org/10.22323/1.305.0015}{\emph{PoS} {\bfseries TASI2017}
  (2017) 015} [\href{https://arxiv.org/abs/1711.00864}{{\ttfamily
  1711.00864}}].

\bibitem{Palti:2019pca}
E.~Palti, \emph{{The Swampland: Introduction and Review}},
  \href{https://doi.org/10.1002/prop.201900037}{\emph{Fortsch. Phys.}
  {\bfseries 67} (2019) 1900037}
  [\href{https://arxiv.org/abs/1903.06239}{{\ttfamily 1903.06239}}].

\bibitem{vanBeest:2021lhn}
M.~van Beest, J.~Calder\'on-Infante, D.~Mirfendereski and I.~Valenzuela,
  \emph{{Lectures on the Swampland Program in String Compactifications}},
  \href{https://arxiv.org/abs/2102.01111}{{\ttfamily 2102.01111}}.

\bibitem{Grana:2021zvf}
M.~Gra\~na and A.~Herr\'aez, \emph{{The Swampland Conjectures: A Bridge from
  Quantum Gravity to Particle Physics}},
  \href{https://doi.org/10.3390/universe7080273}{\emph{Universe} {\bfseries 7}
  (2021) 273} [\href{https://arxiv.org/abs/2107.00087}{{\ttfamily
  2107.00087}}].

\bibitem{Ooguri:2016pdq}
H.~Ooguri and C.~Vafa, \emph{{Non-supersymmetric AdS and the Swampland}},
  \href{https://doi.org/10.4310/ATMP.2017.v21.n7.a8}{\emph{Adv. Theor. Math.
  Phys.} {\bfseries 21} (2017) 1787}
  [\href{https://arxiv.org/abs/1610.01533}{{\ttfamily 1610.01533}}].

\bibitem{Freivogel:2016qwc}
B.~Freivogel and M.~Kleban, \emph{{Vacua Morghulis}},
  \href{https://arxiv.org/abs/1610.04564}{{\ttfamily 1610.04564}}.

\bibitem{Maldacena:1998uz}
J.~M. Maldacena, J.~Michelson and A.~Strominger, \emph{{Anti-de Sitter
  fragmentation}},
  \href{https://doi.org/10.1088/1126-6708/1999/02/011}{\emph{JHEP} {\bfseries
  02} (1999) 011} [\href{https://arxiv.org/abs/hep-th/9812073}{{\ttfamily
  hep-th/9812073}}].

\bibitem{Gaiotto:2009mv}
D.~Gaiotto and A.~Tomasiello, \emph{{The gauge dual of Romans mass}},
  \href{https://doi.org/10.1007/JHEP01(2010)015}{\emph{JHEP} {\bfseries 01}
  (2010) 015} [\href{https://arxiv.org/abs/0901.0969}{{\ttfamily 0901.0969}}].

\bibitem{Antonelli:2019nar}
R.~Antonelli and I.~Basile, \emph{{Brane annihilation in non-supersymmetric
  strings}}, \href{https://doi.org/10.1007/JHEP11(2019)021}{\emph{JHEP}
  {\bfseries 11} (2019) 021}
  [\href{https://arxiv.org/abs/1908.04352}{{\ttfamily 1908.04352}}].

\bibitem{Apruzzi:2019ecr}
F.~Apruzzi, G.~Bruno De~Luca, A.~Gnecchi, G.~Lo~Monaco and A.~Tomasiello,
  \emph{{On AdS$_{7}$ stability}},
  \href{https://doi.org/10.1007/JHEP07(2020)033}{\emph{JHEP} {\bfseries 07}
  (2020) 033} [\href{https://arxiv.org/abs/1912.13491}{{\ttfamily
  1912.13491}}].

\bibitem{Bena:2020xxb}
I.~Bena, K.~Pilch and N.~P. Warner, \emph{{Brane-Jet Instabilities}},
  \href{https://doi.org/10.1007/JHEP10(2020)091}{\emph{JHEP} {\bfseries 10}
  (2020) 091} [\href{https://arxiv.org/abs/2003.02851}{{\ttfamily
  2003.02851}}].

\bibitem{Suh:2020rma}
M.~Suh, \emph{{The non-SUSY AdS$_{6}$ and AdS$_{7}$ fixed points are brane-jet
  unstable}}, \href{https://doi.org/10.1007/JHEP10(2020)010}{\emph{JHEP}
  {\bfseries 10} (2020) 010}
  [\href{https://arxiv.org/abs/2004.06823}{{\ttfamily 2004.06823}}].

\bibitem{Guarino:2020jwv}
A.~Guarino, J.~Tarrio and O.~Varela, \emph{{Brane-jet stability of
  non-supersymmetric AdS vacua}},
  \href{https://doi.org/10.1007/JHEP09(2020)110}{\emph{JHEP} {\bfseries 09}
  (2020) 110} [\href{https://arxiv.org/abs/2005.07072}{{\ttfamily
  2005.07072}}].

\bibitem{Guarino:2020flh}
A.~Guarino, E.~Malek and H.~Samtleben, \emph{{Stable Nonsupersymmetric
  Anti\textendash{}de Sitter Vacua of Massive IIA Supergravity}},
  \href{https://doi.org/10.1103/PhysRevLett.126.061601}{\emph{Phys. Rev. Lett.}
  {\bfseries 126} (2021) 061601}
  [\href{https://arxiv.org/abs/2011.06600}{{\ttfamily 2011.06600}}].

\bibitem{Basile:2021vxh}
I.~Basile, \emph{{Supersymmetry Breaking and Stability in String Vacua: brane
  dynamics, bubbles and the swampland}},
  \href{https://doi.org/10.1007/s40766-021-00024-9}{\emph{Riv. Nuovo Cim.}
  {\bfseries 1} (2021) 98} [\href{https://arxiv.org/abs/2107.02814}{{\ttfamily
  2107.02814}}].

\bibitem{Apruzzi:2021nle}
F.~Apruzzi, G.~Bruno De~Luca, G.~Lo~Monaco and C.~F. Uhlemann,
  \emph{{Non-supersymmetric AdS$_6$ and the swampland}},
  \href{https://arxiv.org/abs/2110.03003}{{\ttfamily 2110.03003}}.

\bibitem{Bomans:2021ara}
P.~Bomans, D.~Cassani, G.~Dibitetto and N.~Petri, \emph{{Bubble instability of
  mIIA on $\mathrm{AdS}_4\times S^6$}},
  \href{https://arxiv.org/abs/2110.08276}{{\ttfamily 2110.08276}}.

\bibitem{Villadoro:2005cu}
G.~Villadoro and F.~Zwirner, \emph{{N=1 effective potential from dual type-IIA
  D6/O6 orientifolds with general fluxes}},
  \href{https://doi.org/10.1088/1126-6708/2005/06/047}{\emph{JHEP} {\bfseries
  06} (2005) 047} [\href{https://arxiv.org/abs/hep-th/0503169}{{\ttfamily
  hep-th/0503169}}].

\bibitem{DeWolfe:2005uu}
O.~DeWolfe, A.~Giryavets, S.~Kachru and W.~Taylor, \emph{{Type IIA moduli
  stabilization}},
  \href{https://doi.org/10.1088/1126-6708/2005/07/066}{\emph{JHEP} {\bfseries
  07} (2005) 066} [\href{https://arxiv.org/abs/hep-th/0505160}{{\ttfamily
  hep-th/0505160}}].

\bibitem{Camara:2005dc}
P.~G. Camara, A.~Font and L.~E. Ibanez, \emph{{Fluxes, moduli fixing and
  MSSM-like vacua in a simple IIA orientifold}},
  \href{https://doi.org/10.1088/1126-6708/2005/09/013}{\emph{JHEP} {\bfseries
  09} (2005) 013} [\href{https://arxiv.org/abs/hep-th/0506066}{{\ttfamily
  hep-th/0506066}}].

\bibitem{Narayan:2010em}
P.~Narayan and S.~P. Trivedi, \emph{{On The Stability Of Non-Supersymmetric AdS
  Vacua}}, \href{https://doi.org/10.1007/JHEP07(2010)089}{\emph{JHEP}
  {\bfseries 07} (2010) 089} [\href{https://arxiv.org/abs/1002.4498}{{\ttfamily
  1002.4498}}].

\bibitem{Lust:2019zwm}
D.~L{\"u}st, E.~Palti and C.~Vafa, \emph{{AdS and the Swampland}},
  \href{https://arxiv.org/abs/1906.05225}{{\ttfamily 1906.05225}}.

\bibitem{Buratti:2020kda}
G.~Buratti, J.~Calderon, A.~Mininno and A.~M. Uranga, \emph{{Discrete
  Symmetries, Weak Coupling Conjecture and Scale Separation in AdS Vacua}},
  \href{https://doi.org/10.1007/JHEP06(2020)083}{\emph{JHEP} {\bfseries 06}
  (2020) 083} [\href{https://arxiv.org/abs/2003.09740}{{\ttfamily
  2003.09740}}].

\bibitem{Acharya:2006ne}
B.~S. Acharya, F.~Benini and R.~Valandro, \emph{{Fixing moduli in exact type
  IIA flux vacua}},
  \href{https://doi.org/10.1088/1126-6708/2007/02/018}{\emph{JHEP} {\bfseries
  02} (2007) 018} [\href{https://arxiv.org/abs/hep-th/0607223}{{\ttfamily
  hep-th/0607223}}].

\bibitem{Saracco:2012wc}
F.~Saracco and A.~Tomasiello, \emph{{Localized O6-plane solutions with Romans
  mass}}, \href{https://doi.org/10.1007/JHEP07(2012)077}{\emph{JHEP} {\bfseries
  07} (2012) 077} [\href{https://arxiv.org/abs/1201.5378}{{\ttfamily
  1201.5378}}].

\bibitem{Junghans:2020acz}
D.~Junghans, \emph{{O-Plane Backreaction and Scale Separation in Type IIA Flux
  Vacua}}, \href{https://doi.org/10.1002/prop.202000040}{\emph{Fortsch. Phys.}
  {\bfseries 68} (2020) 2000040}
  [\href{https://arxiv.org/abs/2003.06274}{{\ttfamily 2003.06274}}].

\bibitem{Marchesano:2020qvg}
F.~Marchesano, E.~Palti, J.~Quirant and A.~Tomasiello, \emph{{On supersymmetric
  AdS$_{4}$ orientifold vacua}},
  \href{https://doi.org/10.1007/JHEP08(2020)087}{\emph{JHEP} {\bfseries 08}
  (2020) 087} [\href{https://arxiv.org/abs/2003.13578}{{\ttfamily
  2003.13578}}].

\bibitem{Cribiori:2021djm}
N.~Cribiori, D.~Junghans, V.~Van~Hemelryck, T.~Van~Riet and T.~Wrase,
  \emph{{Scale-separated AdS$_4$ vacua of IIA orientifolds and M-theory}},
  \href{https://arxiv.org/abs/2107.00019}{{\ttfamily 2107.00019}}.

\bibitem{Marchesano:2019hfb}
F.~Marchesano and J.~Quirant, \emph{{A Landscape of AdS Flux Vacua}},
  \href{https://doi.org/10.1007/JHEP12(2019)110}{\emph{JHEP} {\bfseries 12}
  (2019) 110} [\href{https://arxiv.org/abs/1908.11386}{{\ttfamily
  1908.11386}}].

\bibitem{Lanza:2019xxg}
S.~Lanza, F.~Marchesano, L.~Martucci and D.~Sorokin, \emph{{How many fluxes fit
  in an EFT?}}, \href{https://doi.org/10.1007/JHEP10(2019)110}{\emph{JHEP}
  {\bfseries 10} (2019) 110}
  [\href{https://arxiv.org/abs/1907.11256}{{\ttfamily 1907.11256}}].

\bibitem{Lanza:2020qmt}
S.~Lanza, F.~Marchesano, L.~Martucci and I.~Valenzuela, \emph{{Swampland
  Conjectures for Strings and Membranes}},
  \href{https://arxiv.org/abs/2006.15154}{{\ttfamily 2006.15154}}.

\bibitem{Aharony:2008wz}
O.~Aharony, Y.~E. Antebi and M.~Berkooz, \emph{{On the Conformal Field Theory
  Duals of type IIA AdS(4) Flux Compactifications}},
  \href{https://doi.org/10.1088/1126-6708/2008/02/093}{\emph{JHEP} {\bfseries
  02} (2008) 093} [\href{https://arxiv.org/abs/0801.3326}{{\ttfamily
  0801.3326}}].

\bibitem{Koerber:2007jb}
P.~Koerber and L.~Martucci, \emph{{D-branes on AdS flux compactifications}},
  \href{https://doi.org/10.1088/1126-6708/2008/01/047}{\emph{JHEP} {\bfseries
  01} (2008) 047} [\href{https://arxiv.org/abs/0710.5530}{{\ttfamily
  0710.5530}}].

\bibitem{Ibanez:2012zz}
L.~E. Ibanez and A.~M. Uranga, \emph{{String theory and particle physics: An
  introduction to string phenomenology}}. Cambridge University Press, 2, 2012.

\bibitem{Font:2006na}
A.~Font, L.~E. Ibanez and F.~Marchesano, \emph{{Coisotropic D8-branes and
  model-building}},
  \href{https://doi.org/10.1088/1126-6708/2006/09/080}{\emph{JHEP} {\bfseries
  09} (2006) 080} [\href{https://arxiv.org/abs/hep-th/0607219}{{\ttfamily
  hep-th/0607219}}].

\bibitem{Bergshoeff:2001pv}
E.~Bergshoeff, R.~Kallosh, T.~Ortin, D.~Roest and A.~Van~Proeyen, \emph{{New
  formulations of D = 10 supersymmetry and D8 - O8 domain walls}},
  \href{https://doi.org/10.1088/0264-9381/18/17/303}{\emph{Class. Quant. Grav.}
  {\bfseries 18} (2001) 3359}
  [\href{https://arxiv.org/abs/hep-th/0103233}{{\ttfamily hep-th/0103233}}].

\bibitem{Grimm:2004ua}
T.~W. Grimm and J.~Louis, \emph{{The Effective action of type IIA Calabi-Yau
  orientifolds}},
  \href{https://doi.org/10.1016/j.nuclphysb.2005.04.007}{\emph{Nucl. Phys. B}
  {\bfseries 718} (2005) 153}
  [\href{https://arxiv.org/abs/hep-th/0412277}{{\ttfamily hep-th/0412277}}].

\bibitem{Marchesano:2014iea}
F.~Marchesano, D.~Regalado and G.~Zoccarato, \emph{{On D-brane moduli
  stabilisation}}, \href{https://doi.org/10.1007/JHEP11(2014)097}{\emph{JHEP}
  {\bfseries 11} (2014) 097} [\href{https://arxiv.org/abs/1410.0209}{{\ttfamily
  1410.0209}}].

\bibitem{Carta:2016ynn}
F.~Carta, F.~Marchesano, W.~Staessens and G.~Zoccarato, \emph{{Open string
  multi-branched and K\"ahler potentials}},
  \href{https://doi.org/10.1007/JHEP09(2016)062}{\emph{JHEP} {\bfseries 09}
  (2016) 062} [\href{https://arxiv.org/abs/1606.00508}{{\ttfamily
  1606.00508}}].

\bibitem{Palti:2008mg}
E.~Palti, G.~Tasinato and J.~Ward, \emph{{WEAKLY-coupled IIA Flux
  Compactifications}},
  \href{https://doi.org/10.1088/1126-6708/2008/06/084}{\emph{JHEP} {\bfseries
  06} (2008) 084} [\href{https://arxiv.org/abs/0804.1248}{{\ttfamily
  0804.1248}}].

\bibitem{Escobar:2018rna}
D.~Escobar, F.~Marchesano and W.~Staessens, \emph{{Type IIA flux vacua and
  $\alpha'$-corrections}},
  \href{https://doi.org/10.1007/JHEP06(2019)129}{\emph{JHEP} {\bfseries 06}
  (2019) 129} [\href{https://arxiv.org/abs/1812.08735}{{\ttfamily
  1812.08735}}].

\bibitem{Miyaoka1987}
Y.~Miyaoka, \emph{The chern class and kodaira dimension of a minimal variety.},
  {\emph{Adv. Stud. Pure Math.} {\bfseries 10} (1987) 449}.

\bibitem{DeLuca:2021mcj}
G.~B. De~Luca and A.~Tomasiello, \emph{{Leaps and bounds towards scale
  separation}},  \href{https://arxiv.org/abs/2104.12773}{{\ttfamily
  2104.12773}}.

\bibitem{Gibbons:1997xz}
G.~W. Gibbons, \emph{{Born-Infeld particles and Dirichlet p-branes}},
  \href{https://doi.org/10.1016/S0550-3213(97)00795-5}{\emph{Nucl. Phys. B}
  {\bfseries 514} (1998) 603}
  [\href{https://arxiv.org/abs/hep-th/9709027}{{\ttfamily hep-th/9709027}}].

\bibitem{Casas:2022mnz}
G.~F. Casas, F.~Marchesano and D.~Prieto, \emph{{Membranes in AdS4 orientifold
  vacua and their Weak Gravity Conjecture}},
  \href{https://arxiv.org/abs/2204.11892}{{\ttfamily 2204.11892}}.

\bibitem{Evslin:2007ti}
J.~Evslin and L.~Martucci, \emph{{D-brane networks in flux vacua, generalized
  cycles and calibrations}},
  \href{https://doi.org/10.1088/1126-6708/2007/07/040}{\emph{JHEP} {\bfseries
  07} (2007) 040} [\href{https://arxiv.org/abs/hep-th/0703129}{{\ttfamily
  hep-th/0703129}}].

\bibitem{Donaldson:1996kp}
S.~K. Donaldson and R.~P. Thomas, \emph{{Gauge theory in higher dimensions}},
  in \emph{{Conference on Geometric Issues in Foundations of Science in honor
  of Sir Roger Penrose's 65th Birthday}}, pp.~31--47, 6, 1996.

\bibitem{Lust:2008zd}
D.~Lust, F.~Marchesano, L.~Martucci and D.~Tsimpis, \emph{{Generalized
  non-supersymmetric flux vacua}},
  \href{https://doi.org/10.1088/1126-6708/2008/11/021}{\emph{JHEP} {\bfseries
  11} (2008) 021} [\href{https://arxiv.org/abs/0807.4540}{{\ttfamily
  0807.4540}}].

\bibitem{Held:2010az}
J.~Held, D.~Lust, F.~Marchesano and L.~Martucci, \emph{{DWSB in heterotic flux
  compactifications}},
  \href{https://doi.org/10.1007/JHEP06(2010)090}{\emph{JHEP} {\bfseries 06}
  (2010) 090} [\href{https://arxiv.org/abs/1004.0867}{{\ttfamily 1004.0867}}].

\bibitem{Ihl:2006pp}
M.~Ihl and T.~Wrase, \emph{{Towards a Realistic Type IIA T**6/Z(4) Orientifold
  Model with Background Fluxes. Part 1. Moduli Stabilization}},
  \href{https://doi.org/10.1088/1126-6708/2006/07/027}{\emph{JHEP} {\bfseries
  07} (2006) 027} [\href{https://arxiv.org/abs/hep-th/0604087}{{\ttfamily
  hep-th/0604087}}].

\bibitem{Banks:2006hg}
T.~Banks and K.~van~den Broek, \emph{{Massive IIA flux compactifications and
  U-dualities}},
  \href{https://doi.org/10.1088/1126-6708/2007/03/068}{\emph{JHEP} {\bfseries
  03} (2007) 068} [\href{https://arxiv.org/abs/hep-th/0611185}{{\ttfamily
  hep-th/0611185}}].

\bibitem{Escobar:2018tiu}
D.~Escobar, F.~Marchesano and W.~Staessens, \emph{{Type IIA Flux Vacua with
  Mobile D6-branes}},
  \href{https://doi.org/10.1007/JHEP01(2019)096}{\emph{JHEP} {\bfseries 01}
  (2019) 096} [\href{https://arxiv.org/abs/1811.09282}{{\ttfamily
  1811.09282}}].

\bibitem{Marchesano:2020uqz}
F.~Marchesano, D.~Prieto, J.~Quirant and P.~Shukla, \emph{{Systematics of Type
  IIA moduli stabilisation}},
  \href{https://doi.org/10.1007/JHEP11(2020)113}{\emph{JHEP} {\bfseries 11}
  (2020) 113} [\href{https://arxiv.org/abs/2007.00672}{{\ttfamily
  2007.00672}}].

\bibitem{Taylor:1999ii}
T.~R. Taylor and C.~Vafa, \emph{{R R flux on Calabi-Yau and partial
  supersymmetry breaking}},
  \href{https://doi.org/10.1016/S0370-2693(00)00005-8}{\emph{Phys. Lett. B}
  {\bfseries 474} (2000) 130}
  [\href{https://arxiv.org/abs/hep-th/9912152}{{\ttfamily hep-th/9912152}}].

\bibitem{Bielleman:2015ina}
S.~Bielleman, L.~E. Ibanez and I.~Valenzuela, \emph{{Minkowski 3-forms, Flux
  String Vacua, Axion Stability and Naturalness}},
  \href{https://doi.org/10.1007/JHEP12(2015)119}{\emph{JHEP} {\bfseries 12}
  (2015) 119} [\href{https://arxiv.org/abs/1507.06793}{{\ttfamily
  1507.06793}}].

\bibitem{Herraez:2018vae}
A.~Herraez, L.~E. Ibanez, F.~Marchesano and G.~Zoccarato, \emph{{The Type IIA
  Flux Potential, 4-forms and Freed-Witten anomalies}},
  \href{https://doi.org/10.1007/JHEP09(2018)018}{\emph{JHEP} {\bfseries 09}
  (2018) 018} [\href{https://arxiv.org/abs/1802.05771}{{\ttfamily
  1802.05771}}].

\bibitem{BerasaluceGonzalez:2012zn}
M.~Berasaluce-Gonzalez, P.~G. Camara, F.~Marchesano and A.~M. Uranga, \emph{{Zp
  charged branes in flux compactifications}},
  \href{https://doi.org/10.1007/JHEP04(2013)138}{\emph{JHEP} {\bfseries 04}
  (2013) 138} [\href{https://arxiv.org/abs/1211.5317}{{\ttfamily 1211.5317}}].

\bibitem{Marchesano:2014mla}
F.~Marchesano, G.~Shiu and A.~M. Uranga, \emph{{F-term Axion Monodromy
  Inflation}}, \href{https://doi.org/10.1007/JHEP09(2014)184}{\emph{JHEP}
  {\bfseries 09} (2014) 184} [\href{https://arxiv.org/abs/1404.3040}{{\ttfamily
  1404.3040}}].

\bibitem{Blumenhagen:2014gta}
R.~Blumenhagen and E.~Plauschinn, \emph{{Towards Universal Axion Inflation and
  Reheating in String Theory}},
  \href{https://doi.org/10.1016/j.physletb.2014.08.007}{\emph{Phys. Lett. B}
  {\bfseries 736} (2014) 482}
  [\href{https://arxiv.org/abs/1404.3542}{{\ttfamily 1404.3542}}].

\bibitem{Hitchin:1999fh}
N.~J. Hitchin, \emph{{Lectures on special Lagrangian submanifolds}},
  {\emph{AMS/IP Stud. Adv. Math.} {\bfseries 23} (2001) 151}
  [\href{https://arxiv.org/abs/math/9907034}{{\ttfamily math/9907034}}].

\bibitem{Baulieu:1997jx}
L.~Baulieu, H.~Kanno and I.~M. Singer, \emph{{Special quantum field theories in
  eight-dimensions and other dimensions}},
  \href{https://doi.org/10.1007/s002200050353}{\emph{Commun. Math. Phys.}
  {\bfseries 194} (1998) 149}
  [\href{https://arxiv.org/abs/hep-th/9704167}{{\ttfamily hep-th/9704167}}].

\end{thebibliography}\endgroup

\end{document}